\documentclass{kapproc} 
\usepackage{procps} 

\usepackage[dvips]{graphicx}

\upperandlowercase

\sectionequationnumber 

\normallatexbib

\newcommand{\be}{\begin{eqnarray}}
\newcommand{\ee}{\end{eqnarray}}
\newcommand\del{\partial}
\newcommand{\mat}{\left ( \begin{array}{cc}}
\newcommand{\emat}{\end{array} \right )}
\newcommand{\matt}{\left ( \begin{array}{ccc}}
\newcommand{\ematt}{\end{array} \right )}
\newcommand{\matf}{\left ( \begin{array}{cccc}}
\newcommand{\ematf}{\end{array} \right )}
\newcommand{\vect}{\left ( \begin{array}{c}}
\newcommand{\evect}{\end{array} \right )}
\newcommand{\Tr} {\rm Tr}
\newcommand{\nn} {\nonumber}

\begin{document}

\articletitle{QCD, Chiral Random Matrix Theory and Integrability}


\author{J.J.M. Verbaarschot}
\affil{Department of Physics and Astronomy, SUNY at Stony Brook, NY 11794}
\email{jacobus.verbaarschot@stonybrook.edu}





\section{Summary}

Random Matrix Theory has been a unifying approach in physics and mathematics. 
In these lectures we discuss applications of Random Matrix Theory
to QCD and emphasize underlying integrable structures.
In the first lecture we give an overview
of QCD, its low-energy limit and the microscopic limit of the
Dirac spectrum which, as
we will see in the second lecture, can be described by chiral
Random Matrix Theory. 
The main topic of the third lecture is the  recent developments 
on the relation
between the QCD partition function and integrable hierarchies 
(in our case the Toda lattice hierarchy).
This is an efficient way 
to obtain the QCD Dirac spectrum from the low energy limit
of the QCD partition function. Finally, we will discuss the 
QCD Dirac spectrum at nonzero chemical potential. 
We will show that the microscopic spectral density is given
by the replica limit of the Toda lattice equation.
Recent results by Osborn on the Dirac spectrum of
 full QCD will be discussed.




\newpage
\tableofcontents
\section{Introduction}

Applications of Random Matrix Theories to problems in physics have a long
history starting with the idea of Wigner \cite{Wigner} to describe the
spacing distribution of nuclear levels by an ensemble of real symmetric
matrices. Although this is the first application of Random Matrix Theory (RMT)
to strong interactions, applications of RMT to QCD
started much later. The first paper that put QCD into the context
of RMT was the work of 't Hooft on two-dimensional QCD in the
limit of a large number of colors \cite{hooft}. It was  shown
 \cite{brezin} that 
the combinatorial factors that enter in this 
large $N_c$ expansion could be obtained from  matrix integrals. 
Even today, as we have seen in the lectures by Di Francesco 
\cite{francesco}, this
work attracts a great deal of attention. 
It greatly stimulated the analysis of a complicated nonlinear theory
such as QCD
by means of much simpler matrix models. Because of the success of the
application of RMT to the planar expansion of QCD, the hope was that nontrivial
results could be derived this way. I will mention three well-known results
that have emerged from this line of thought: 
the Brezin-Gross-Witten model \cite{gross-witten,brezin-gross},
the Eguchi-Kawai \cite{eguchi}
reduction, and induced QCD according to Kazakov and Migdal 
\cite{kazakov-migdal}.

The Wilson loop in lattice QCD {\em without} quarks in 1+1 dimensions can
be reduced to the calculation of the  unitary matrix integral
\be
z(g^2,N_c) =\int_{U\in SU(N_c)} dU e^{g^{-2} {\rm Tr}(U+ U^\dagger)}.
\label{one-link}
\ee
which is known as the Brezin-Gross-Witten model
\cite{bars-green,gross-witten,brezin-gross}. 
  This reduction was generalized
to an arbitrary Wilson loop amplitude in the large $N_c$ limit of
lattice QCD in four dimensions and is known as the Eguchi-Kawai reduction 
\cite{eguchi}. It was shown that Wilson loop amplitudes
do not depend on space-time and can be obtained from a single plaquette
integral. However, this reduction is not valid in the
weak coupling limit \cite{bhanot-heller-neuberger}. The idea of induced
QCD \cite{kazakov-migdal} is to induce the plaquette action by a unitary
matrix integral.  With a vanishing Wilson line \cite{kogan} 
this approach turned out to be unsuccessful as well.

The matrix model (\ref{one-link}) 
also  appears in the low-energy limit of QCD {\em with} 
quarks. However,
in this case the integral over $SU(N_c)$ is not over the color degrees of
freedom but rather over the flavor degrees of freedom, and $g^{-2}$ is
replaced by  $mV \Sigma/2$ with $\Sigma$ the chiral condensate, $m$
the quark mass and $V$ the volume of space time. It coincides with
the full QCD partition function in a domain where the pion Compton wavelength
is much larger than the size of the box \cite{GaL,LS}. In this limit we have
\be
\frac 1{VN_f} \del_m\, z(m\Sigma V/2,N_f) = \langle \frac 1V 
\sum_k \frac1{i\lambda_k+m} \prod_k(i\lambda_k+m)^{N_f}\rangle,
\label{condin}
\ee
where the $\lambda_k$ are the eigenvalues of the Dirac operator.
By expanding in powers of the inverse mass, 
one obtains sum rules for the inverse Dirac eigenvalues  
\cite{LS} which put constraints on the Dirac spectrum,
but do not determine the average spectral density on the scale 
of the average level spacing (which is known as the microscopic spectral
density) and other spectral correlators. 

A Random Matrix Theory that describes the fluctuations of the small 
eigenvalues of the Dirac operator was introduced in \cite{SV,V}.
It was shown that chiral RMT is equivalent to 
the flavor unitary matrix integral (\ref{one-link}). 
The spectral correlation functions
were found to be in good agreement with lattice QCD 
simulations (see \cite{anrev,poul} for a review of lattice results).
One argument to understand this is that level correlations of
complex systems on the scale of the average level spacing are universal, i.e.
they do not depend on the details of the system. This could be shown 
rigorously in the context of RMT \cite{akedammagnis}.
However, it was understood later that the 
generating function for the Dirac spectrum
is completely determined by chiral symmetry
\cite{OTV,DOTV}, and its microscopic limit does not change
if the Dirac operator is replaced
by a random matrix. This is one of the main topics of these lectures.

The question has been raised if the quenched spectral density can
be obtained from the limit $N_f \to 0$ of (\ref{condin}). This 
procedure is known as the replica trick. It has been argued that
this limit generally gives the wrong result \cite{critique}.
However, the family of partition functions for different values
of $N_f$ are related by the
Toda lattice equation \cite{mmm}. If we take the replica limit of the
Toda lattice equation \cite{SplitVerb1}, or of the corresponding
Painlev$\acute{\rm e}$ equation \cite{kanzieper},
we obtain the correct nonperturbative result.
This is a second main topic of these lectures.

A third main topic of these lectures 
is the discussion of QCD at  nonzero 
baryon chemical
potential. In that case the Dirac operator is non-Hermitian and its
eigenvalues are scattered in the complex plane. 
We will show that also in this case the spectral density can be
obtained from the Toda lattice equation \cite{SplitVerb2,AOSV}.
Recent analytical results by Osborn \cite{james} for the Dirac
spectrum of QCD with dynamical quarks
at nonzero chemical potential will be discussed.

We start these lectures with an elementary introduction to QCD and
its symmetries. The Dirac spectrum is discussed in section \ref{dirac}.
The low energy limit of QCD or partially quenched 
QCD (see section \ref{low}) 
is equivalent to a RMT with the symmetries of
QCD that will be introduced in section \ref{chrmt}. In section \ref{chrmt}
we also calculate the microscopic spectral density
by means of orthogonal polynomials and the supersymmetric method.
In section \ref{integ} we show that this spectral density can be 
obtained from the replica limit of the Toda lattice equation.
In the same section we connect these results with ideas from
the theory of exactly solvable systems such as Virasoro constraints,
Painlev$\acute {\rm e}$ equations, Backlund transformations and 
the Toda lattice. 
QCD at nonzero chemical potential is discussed in sections \ref{qcmu} and
\ref{full}. In section \ref{qcmu} we show that the microscopic spectral 
density can be obtained from the replica limit of a Toda lattice
equation. Recent results for full QCD at nonzero chemical potential
are discussed in section \ref{full} and concluding remarks are made in
section \ref{conclu}. 

Finally, a note about books and reviews on the subject of these lectures.
The classic RMT
text is the book by Mehta 
\cite{Mehta} which emphasizes the orthogonal
polynomial method. The third edition of this book appeared recently.
In the book by Forrester \cite{ForresterBook} the emphasis is
on the relation between RMT and solvable models and mathematical physics.
A comprehensive review of RMT is given in \cite{Guhr98}. 
A collection of recent papers on RMT can be found in 
\cite{special} which also contains several reviews.
Applications to mesoscopic physics are discussed in 
\cite{EfetovBook,Beenreview},
applications to QCD in \cite{anrev} and applications to Quantum Gravity
in \cite{zinn}. Among the pedagogical reviews we mention 
a general review of RMT \cite{ieee} and 
an introduction to the supersymmetric method \cite{fyodorov,mex}.
These lecture notes overlap in part with lecture notes I prepared for the
Latin American Summer School in Mexico City \cite{mex}, where 
the main emphasis was on the supersymmetric method rather 
than on applications of
RMT to QCD. 

\section{QCD}
\label{qcd}
\subsection{Introduction}
QCD (Quantum Chromo Dynamics) is the theory of strong interactions that
describes the world of nucleons, pions and the nuclear force.
No experimental deviations from this theory have ever been detected.
QCD is a theory of quarks which interact via gauge bosons known as
gluons. In nature we have 6 different flavors of quarks which each 
occur in three colors. Each quark is represented by a 4-component
Dirac spinor
\be
q_{i,\,\mu}^f,\qquad f = 1, \cdots, N_f = 6, \qquad i = 1, \cdots, N_c =3
\ee
with Dirac index $\mu$. In total we have 18 quarks (plus an equal number
of anti-quarks).
The gluon fields are represented by the gauge potentials
\be
A_\mu^{ij}, \qquad i, j = 1, \cdots , N_c =3,
\ee
which, as is the case in electrodynamics, are spin 1 vector fields. 
The gauge fields are Hermitian and traceless; they
span the algebra of $SU(N_c)$. In total we have 8 massless gluons.
\begin{table}[!t]
\caption{Quark Masses}
\centering
\begin{tabular}{ll}
\hline
$m_u$ = 4 MeV & $m_c$ = 1.3 GeV\\   
$m_d$ = 8 MeV & $m_b$ = 4.4 GeV \\
$m_s$ = 160 MeV & $m_t$ = 175 GeV
\end{tabular}
\end{table}
The 6 quark flavors are known as up, down, strange, charm, bottom and
and top. The quark masses are given in Table 1.
Only the two lightest quarks are important for low-energy nuclear physics.

First principle calculations of QCD can be divided into
three different groups, perturbative QCD, lattice QCD and chiral
perturbation theory. The main domain of applicability of 
perturbative QCD is for momenta above several GeV. Chiral perturbation
theory is an expansion in powers of the momentum and the pion mass
and is only reliable below several hundred MeV. Although lattice QCD
is an exact reformulation of QCD, in practice both the domain of low
momenta and high momenta cannot be accessed, and its main domain of
applicability lies somewhere in between the two perturbative schemes.

The reason that perturbative QCD is applicable at high energies is
asymptotic freedom: the coupling constant $g \to 0$ for 
momenta $p \to \infty$. This property was instrumental
in gaining broad acceptance for QCD as the theory of strong
interactions and its discovers were awarded this years Nobel prize.

A second important property of QCD is confinement, meaning that
only color singlets can exist as asymptotic states. This
empirically known property has been confirmed by lattice QCD calculations.
However, a first principle proof of the existence of a mass gap in
QCD is still lacking even in the absence of quarks.  
Because of confinement the lightest particles of the theory are not
quarks or gluons but rather composite mesons.

This brings us to the third important property of QCD: chiral symmetry.
At low temperatures and density chiral symmetry is broken spontaneously 
which,
according to Goldstone's, theorem gives rises to massless Goldstone
bosons. Because the chiral symmetry is slightly broken by the 
light quark masses, the Goldstone bosons are not exactly massless,
but the mass of the pions of 135-138 ${\rm MeV}$ is an order of magnitude
less than a typical QCD scale of about 1 ${\rm GeV}$. 
This justifies a systematic expansion in the pion mass
and the momenta known as chiral
perturbation theory.

A fourth important property of QCD is that a first principle
nonperturbative lattice
formulation can be simulated numerically. This allows us
to compute nonperturbative observables such as for example the nucleon
mass and $\rho$-meson mass. Without lattice QCD we would have had
only a small number of first principle nonperturbative results and
the validity of QCD in this domain would still have been a big question mark.

\subsection{The QCD partition function}
\label{symme1}

The QCD partition function in a
box of volume $V_3=L^3$ can be expressed in terms of the eigenvalues of the
QCD Hamiltonian $E_k$ as
\be
Z^{\rm QCD} = \sum_k e^{-\beta E_k},
\ee
where $\beta$ is the inverse temperature. At low temperatures, $(\beta 
\rightarrow \infty)$, the partition function is dominated by the lightest
states of the theory, namely the vacuum state, with an energy density of 
$E_0/V_3$ and massless excitations thereof. 
The partition function $Z^{\rm QCD}$
can be rewritten as a
Euclidean functional integral over the nonabelian gauge fields $A_\mu$,
\be
Z^{\rm QCD}(M) = \int d A_\mu \prod_{f=1}^{N_f} \det (D + m_f) 
e^{-S^{\rm YM} },
\label{ZQCD}
\ee
where $S^{\rm YM}$ is the Yang-Mills action given by
\be
S^{\rm YM} = \int d^4x [\frac 1{4g^2}{ F_{\mu\nu}^a}^2 
- i\frac {\theta}{32\pi^2}
F_{\mu\nu}^a \tilde F_{\mu\nu}^a].
\label{SYM}
\ee
The field strength and its dual are given by
\be
F_{\mu\nu}^a = \partial_\mu A_{\nu}^a -\partial_\nu A_\mu^a + f_{abc}
A_\mu^b A_\nu^c,\qquad
\tilde F_{\mu\nu}= \frac 12 \epsilon_{\mu\nu\alpha\beta} F^{\alpha \beta}.
\ee
The $f_{abc} $ are the structure constants of the gauge group $SU(N_c)$.
The gauge fields are denoted by
$A_\mu = A_{\mu a} \frac{T^a}{2}$,
where $T^a$ are the generators of  the gauge group. 
The integral $\nu \equiv 
\frac 1{32\pi^2}\int d^4x F_{\mu\nu}^a \tilde F_{\mu\nu}^a$
is a topological invariant, i.e. it does not change under continuous 
transformations of the gauge fields. An important class of field configurations
are instantons. These are topological nontrivial field configurations that
minimize the classical action. They are classified according to their
topological charge $\nu$. The parameter $\theta$ is known
as the $\theta$-angle. Experimentally, its value is consistent with zero.
In (\ref{ZQCD}), the mass matrix is diagonal, $M = {\rm diag}
(m_1,\cdots, m_{N_f})$, but below we will also consider a general mass
matrix. 
The anti-Hermitian Dirac operator in (\ref{ZQCD}) is given by
\be
D = \gamma_\mu(\partial_\mu +i A_\mu),
\ee 
where the $\gamma_\mu$ are the Euclidean Dirac matrices with
anti-commutation relation
$
\{ \gamma_\mu, \gamma_\nu \} = 2 \delta_{\mu\nu}.
$
In the chiral representation the $\gamma$-matrices are given by
\be
\gamma_k = \mat 0 & i\sigma_k \\ -i\sigma_k & 0  \emat, \quad
\gamma_4 = \mat 0 & 1 \\ 1 & 0 \emat, \quad 
\gamma_5 = \mat 1 & 0 \\ 0 & -1 \emat.
\ee
In this representation the Dirac operator has the structure
\be
D = \mat 0 & id \\ id^\dagger & 0 \emat.
\label{diracstructure}
\ee
The integration measure is defined by discretizing space-time
\be
d A_\mu^a = \prod_x dA_\mu^a(x).
\ee 
A particular popular discretization is the lattice discretization where
the QCD action is discretized on a hyper-cubic lattice with spacing
$a$. The discussion of lattice QCD would be a lecture by itself. For
the interested reader we recommended 
several excellent textbooks on the subject \cite{creutz,rothe,montvay}.

A field theory is obtained by taking the continuum limit, i.e. the 
limit of zero lattice spacing $a$   for the integration
measure discussed above. This limit only exists if we simultaneously
adjust the coupling constant, i.e. $ g \rightarrow g(a)$.
If such limit exists the field theory is called renormalizable.
For QCD $g(a)$ approaches
zero in the continuum limit, a property known as asymptotic freedom.

We will be mainly interested in the eigenvalues of the Dirac operator
and how they fluctuate for gauge fields $A_\mu$ distributed according
to the QCD action. We will show that below a well-defined
scale the fluctuations
of the Dirac eigenvalues are given by a RMT with
the global symmetries of the QCD.

\subsection{Symmetries of the QCD Partition Function}
\label{symme}

It is well-known that the QCD action is greatly 
constrained by gauge symmetry, Poincar$\acute{\rm e}$ invariance 
and renormalizability. These symmetries determine the structure
of the Dirac operator and are essential for its infrared spectral
properties. 
In this section we will discuss
the global symmetries of the Euclidean Dirac operator. 
In particular,  
the chiral symmetry, the flavor symmetry and
the anti-unitary symmetry of the continuum Dirac operator are discussed.

\subsubsection{Axial Symmetry}
\label{symme2}

The axial symmetry, or the $U_A(1)$ symmetry, can be expressed as
the anti-commutation relation
\be
\{\gamma_5, D\} =0.
\label{ua1}
\ee
This implies that all nonzero eigenvalues occur in pairs $\pm i\lambda_k$ with
eigenfunctions given  by $\phi_k$ and $\gamma_5 \phi_k$. If $\lambda_k = 0$ the
possibility exists that $\gamma_5 \phi_k \sim \phi_k$, so that $\lambda_k=0$ 
is an unpaired eigenvalue. According to the Atiyah-Singer index theorem, 
the total number of such zero eigenvalues
is a topological invariant, i.e., it does not change under
continuous transformations of the gauge field configuration.
Indeed, this possibility is realized by  the field of an instanton 
which is a solution
of the classical equations of motion. On the other hand, it cannot
be excluded that $\lambda_k = 0$ while $\phi_k$ and $\gamma_5\phi_k$ are
linearly independent. However, this imposes additional constraints on
the gauge fields that will be violated by infinitesimal deformations.
Generically, such situation does not occur.
 
In a decomposition according to the total number of topological zero modes,
the QCD partition function can be written as
\be
Z^{\rm QCD}(M,\theta) = \sum_{\nu=-\infty}^\infty
 e^{i\nu\theta} Z_\nu^{\rm QCD}(M),
\label{ZQCDdet}
\ee
where
\be
Z^{\rm QCD}_\nu(M) = \langle\prod_f m^{\nu}_f \prod_k 
(\lambda_k^2 + m^2_f)
\rangle_\nu.
\ee
Here, $\langle \cdots \rangle_\nu$ denotes the average over gauge-field
configurations with topological charge $\nu$ weighted by the Yang-Mills
action. If we introduce  right-handed and left-handed masses as complex
conjugated masses we find that the $\theta$ dependence of the QCD
partition function is only through the combination $m e^{i\theta/N_f}$.
This property can be used to obtain the $\theta$-dependence of the
low-energy effective partition function.

\subsubsection{Flavor Symmetry}
\label{symme3}

A second important global symmetry is the flavor symmetry. This symmetry
can be best explained by writing the fermion determinant in the QCD partition
function as a functional  integral over Grassmann variables,
\be
\prod_f \det(D+m_f) = \int d \psi d \bar \psi e^{\int d^4x
\sum_{f=1}^{N_f} \bar \psi^f (D+ m_f) \psi^f}.
\ee
In a chiral basis with $\psi_R = \gamma_5 \psi_R$
and $\psi_L = -\gamma_5 \psi_L$, the exponent can be rewritten as
\be
\sum_{f=1}^{N_f} \bar \psi^f (D+ m_f) \psi^f=
\bar \psi^f_R D \psi^f_R + \bar \psi^f_L D \psi^f_L +
\bar \psi^f_R  M_{RL} \psi^f_L+ \bar \psi^f_L  M_{LR}\psi^f_R.\nonumber \\
\ee

To better illuminate the transformation
properties of the partition function we have replaced the diagonal
mass matrix by $M_{RL}$ and $M_{LR}$.

For $m_f = 0$ we have the symmetry
\be
  \psi_L \rightarrow U_L \psi_L, \quad
\bar  \psi_L \rightarrow \bar \psi_LU^{-1}_L, \quad
\psi_R \rightarrow U_R \psi_R, \quad
\bar \psi_R \rightarrow \bar\psi_R U_R^{-1}.
\label{flavorsym}
\ee
The only condition to be imposed on $U$ and $V$ is that their inverse exists.
If the number of left-handed modes is equal to the number of right-handed modes
we thus have an invariance under $Gl_R(N_f) \times Gl_L(N_f)$, where 
$Gl(N_f)$ is the group of complex $N_f\times N_f$ matrices with nonzero
determinant.  
However, if
the number of left-handed modes is not equal to the number of right-handed
modes, the axial-symmetry group
is broken to an $Sl(N_f)$ subgroup whereas the vector
symmetry with $U_L=U_R$ remains unbroken.
For $m_f=0$ the flavor symmetry is thus broken explicitly to 
$Gl_V(N_f)\times Sl_A(N_f)$ by instantons or the anomaly.
A $Gl_V(1)$ subgroup of $Gl_V(N_f)$ corresponds to baryon number
conservation and is usually not considered when flavor symmetries
are discussed. 

What is much more important, though, is the spontaneous breaking of
the axial flavor symmetry. From lattice QCD simulations and phenomenological
arguments we know that the expectation value
$
\langle \bar \psi \psi \rangle =
\langle \bar \psi_R \psi_L \rangle +
\langle \bar \psi_L \psi_R \rangle\approx -(240\, {\rm MeV})^3
$
in the vacuum state of QCD instead of the symmetric possibility
$\langle \bar \psi \psi \rangle = 0$.
Phenomenologically, this is known 
because the pions are much lighter than
the $\sigma$ mesons. The spontaneous breaking of the axial symmetry  
also follows from the absence of
parity doublets. For example, the pion mass and the $a_0$ mass
are very different ($m_\pi = 135\,{\rm MeV}$ and $m_\delta = 980\,  
{\rm MeV}$).

For fermionic quarks there is no need to extend the
symmetry group to   $Gl_R(N_f) \times Gl_L(N_f)$. In that case, the flavor
symmetry is spontaneously broken according to 
$SU_R(N_f) \times SU_L(N_f)\to SU_V(N_f)$. 
For bosonic quarks, which enter in the generating function
for the Dirac spectrum, it will be shown in the next section that it
is essential to consider the complex
extension of $SU(N_f)$.
Notice that the complex extension of
a symmetry group does not change the number of conserved 
currents. 

On easily verifies  
that $\langle \bar \psi \psi \rangle$ is only invariant for $U_L=U_R$.
The vacuum state thus breaks the chiral symmetry down to $Gl_V(N_f)$.
In agreement
with the Vafa-Witten theorem \cite{Vafa} only the axial symmetries
can be broken spontaneously.
We also observe that the {\em complete}
axial group is broken which is known as the maximum breaking 
\cite{Shifman-three} of chiral symmetry.

\subsubsection{Flavor Symmetry for Bosonic Quarks}

For bosonic quarks the Goldstone bosons cannot be parameterized by 
a unitary matrix. The reason is that symmetry transformations have
to be consistent with the convergence of the bosonic integrals.
Let us consider the case of one bosonic flavor. Then
\be
{\det}^{-1} \mat m & id\\ id^\dagger & m \emat
= \frac 1{\pi^2}\int d^2\phi_1 d^2\phi_2 \exp \left [-
\vect \phi_1^* \\ \phi_2^* \evect \mat m & id\\ id^\dagger & m \emat
\vect \phi_1 \\ \phi_2 \evect \right ],\nn\\ \qquad
\label{detinvbos}
\ee
so that the integral is convergent for ${\rm Re}(m) > 0$.
The most general flavor symmetry
group of the action in (\ref{detinvbos}) 
is $Gl(2)$ that can be parameterized as 
\be
U = e^H V \quad {\rm with } \quad H^\dagger = H \quad {\rm and } \quad
V V^\dagger =1 .
\label{udecompos}
\ee
For $U$ to be a symmetry transformation for $m = 0$  we require that
\be
U^\dagger \mat 0 & id \\ id^\dagger & 0 \emat U =
 \mat 0 & id \\ id^\dagger & 0 \emat,
\ee
so that $H$ has to be a multiple of $\sigma_3$, and $V$ has to be a 
multiple of the identity. 
The transformations $V$ in (\ref{udecompos}) are not broken by  
the mass term and therefore represent the vector symmetry. Only the symmetry
transformation $\exp(s\sigma_3)$ is broken by the mass term
 so that the axial transformations are parameterized by
\be
U = \mat e^s & 0 \\ 0 & e^{-s} \emat \qquad {\rm with} 
\quad s \in \langle -\infty, \infty \rangle.
\ee
For $N_f$ bosonic flavors the axial transformations are parameterized by
\be
U = \mat  e^H & 0 \\ 0 & e^{-H} \emat \qquad{\rm with}\qquad H^\dagger  = H,
\ee
which is the coset $Gl(N_f)/U(N_f)$.

\subsection{ Anti-Unitary Symmetries and the Three-fold Way}
\label{symme4}

The QCD partition function with three or more colors in the fundamental
representations has no anti-unitary symmetries. As will be
discussed below, for two colors with
fundamental fermions and for adjoint fermions, the Dirac operator
has an anti-unitary symmetry. The classification
of the QCD Dirac operator 
according to anti-unitary symmetries was introduced in \cite{V}.

\subsubsection{Anti-Unitary symmetries and the Dyson index}
\label{rando4}

The value of the Dyson index is determined by the anti-unitary symmetries
of the system. If there are no anti-unitary symmetries the Hamiltonian
is Hermitian and the value of the Dyson index is $\beta_D = 2$. 

An anti-unitary symmetry operator, which can always be written as $A= UK$ with
$U$ unitary and $K$ the complex conjugation operator,  commutes
with the Hamiltonian of the system
\be
[H,UK] = 0.
\label{commutator}
\ee 
We can distinguish
two possibilities,
\be
(UK)^2 = 1 \quad {\rm or} \quad (UK)^2 =-1,
\ee
corresponding to $\beta_D =1$ and $\beta_D=4$, respectively.
The argument goes as follows. The 
symmetry operator $A^2=(UK)^2=UU^*$ is unitary, and in an irreducible
subspace, it is necessarily a multiple of the identity, $UU^* = \lambda
{\bf 1}$.  Because of this relation, $U$ and $U^*$ commute so that
$\lambda$ is real. By unitarity we have $|\lambda| = 1$ which yields
$\lambda=\pm1$.

When $\beta_D = 1$ it is always possible to find a basis in which the
Hamiltonian is real. Starting  with basis vector $\phi_1$ we can
construct $\psi_1 =\phi_1 + UK \phi_1$. Then choose $\phi_2$ perpendicular
to $\psi_1$ and define $\psi_2 =\phi_2 + UK \phi_2$ with
\be
(\phi_2+ UK \phi_2, \psi_1) 
= (UK \phi_2,\psi_1) 
= ((UK)^2 \phi_2,UK\psi_1)^*
= (\phi_2,\psi_1)^* = 0.\nn
\ee
The next basis vector is found by choosing $\phi_3$ perpendicular
to $\psi_1$ and $\psi_2$, etc.\,.
In this basis the Hamiltonian is real
\be
H_{kl} &=& (\psi_k, H \psi_l)
= (UK\psi_k, UKH \psi_l)^* 
= (\psi_k, HUK \psi_l)^* 
= (\psi_k,H \psi_l)^*\nn\\  &=&H_{kl}^*.
\label{timerev}
\ee

The best known anti-unitary operator in this class is the time-reversal
operator for which  $U$ is the identity matrix.

In the case $(UK)^2 = -1$ all eigenvalues of the Hamiltonian are
doubly degenerate. This can be shown as follows. If $\phi_k$ is and
eigenvector with eigenvalue $\lambda_k$, then it follows from 
(\ref{commutator}) that also $UK\phi_k$ is an eigenvector of
the Hamiltonian with the same eigenvalue. The important thing is
that this eigenvector is perpendicular to $\phi_k$ \cite{LS},
\be
(\phi_k, UK\phi_k) = (UK \phi_k , (UK)^2 \phi_k)^*= -(\phi_k, UK\phi_k).
\label{independent}
\ee
In this case it is possible to construct a basis for
which the Hamiltonian matrix can be organized into real quaternions
\cite{Porter}. The eigenvalues of a Hermitian quaternion real matrix
are quaternion scalars, and the eigenvalues of the original matrix 
are thus doubly degenerate in agreement with (\ref{independent}).
The best known example in this class is the Kramers degeneracy  for 
time reversal invariant systems with  half-integer spin but no
rotational invariance. For example, for spin $\frac 12$ the
time reversal operator is
given by $\sigma_2 K$ with $(K\sigma_2)^2 = -1$.

Next we will discuss the anti-unitary symmetries
of the QCD Dirac operator.

\subsubsection{QCD in the Fundamental Representation}
\label{symme41}

For three or more colors, QCD in the fundamental representation does
not have any anti-unitary symmetries and $\beta_D =2$. QCD with two
colors is exceptional. The reason is 
the pseudo-reality of $SU(2)$:
\be
A_\mu^* =   (\sum_k A_\mu^k \frac{\tau_k}2)^* =  - \tau_2 A_\mu \tau_2,
\ee
where the $\tau_k$ are the Pauli matrices acting in color space. 
From the explicit representation for the $\gamma$-matrices it follows that
\be
\gamma_\mu^* = \gamma_2 \gamma_4 \gamma_\mu \gamma_2\gamma_4 .
\label{gammac}
\ee
For the Dirac operator $iD = i\gamma_\mu \partial_\mu + \gamma_\mu A_\mu$
we thus have
\be
[KC\gamma_5\tau_2, D] = 0,
\label{anti1}
\ee
where $K$ is the complex conjugation operator and $C= \gamma_2\gamma_4$ is the 
charge conjugation matrix.
Because
$
(KC\gamma_5\tau_2)^2= 1
$
we have that $\beta_D=1$.
Using the argument of Eq. (\ref{timerev})
a basis can be constructed such that the Dirac matrix is real for any $A_\mu$.

\subsubsection{QCD in the Adjoint Representation}
\label{symme42}

For QCD with gauge fields in the adjoint representation the 
Dirac operator is given by
\be
D = \gamma_\mu\partial_\mu\delta_{bc} + f^{abc} \gamma_\mu A_{a\mu},
\ee
where the $f^{abc}$ denote the structure constants of the gauge group. 
Because of the complex conjugation property of the $\gamma$-matrices we 
have that
\be
[\gamma_2\gamma_4 \gamma_5 K,D] = 0.
\ee
One easily verifies that in this case
\be
(\gamma_2\gamma_4 \gamma_5K)^2 = -1,
\ee
so that the eigenvalues of $D$ are doubly degenerate 
(see section \ref{rando4}).
This corresponds to the case $\beta_D = 4$, so that it is possible to 
organize the matrix elements of the Dirac operator into real quaternions.

\section{The Dirac Spectrum in QCD}
\label{dirac}

In this section we show
that the smallest eigenvalues of the QCD Dirac operator are related to
the chiral condensate by means of the Banks-Casher relation. This result is
used to define the microscopic spectral density.
\subsection{Banks-Casher Relation}

The order parameter of the chiral phase transition,
$\langle \bar \psi \psi \rangle$,
is nonzero only below a critical temperature.
As was shown by Banks and Casher {\cite{BC}},
$\langle \bar \psi \psi \rangle$ is directly related to the eigenvalue density
of the QCD Dirac operator per unit four-volume
\be
\Sigma \equiv
|\langle \bar \psi \psi \rangle| =\lim\frac {\pi \langle {\rho(0)}\rangle}V.
\label{bankscasher}
\ee
For eigenvalues
$\{\lambda_k\}$ the average spectral density is given by
\be
\rho(\lambda) = \langle \sum_k \delta(\lambda-\lambda_k) \rangle.
\ee

\begin{figure}[ht] 
\centerline{\includegraphics[height=5cm]{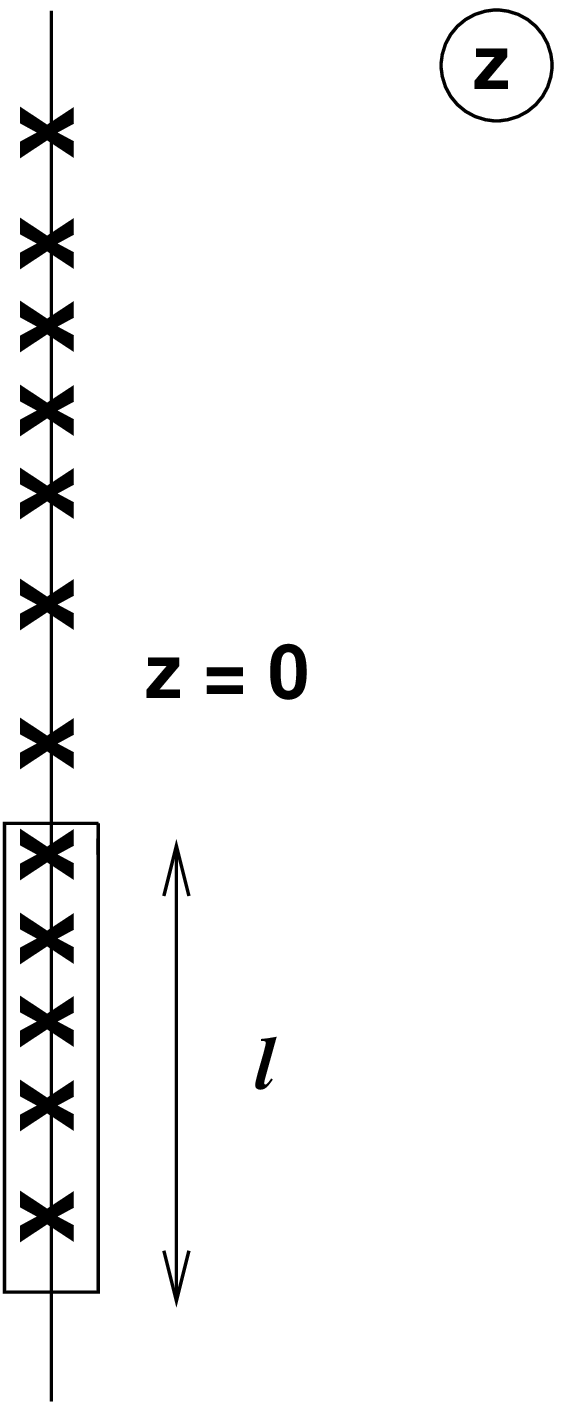}}   
\caption{}{\small A typical Dirac spectrum. To derive the Banks-Casher
relation we integrate the resolvent over the rectangular contour
in this figure. 
(Figure taken from  \cite{minn04}.)}
\label{fig1}
\end{figure}

To show the Banks-Casher relation we study the resolvent defined
by
\be
G(z) = \sum_k \frac 1{z+i\lambda_k}.
\ee
It can be interpreted as the electric field at $z$ of charges
at $i\lambda_k$. Using this analogy it is clear that the resolvent 
changes sign if $z$ crosses the imaginary axis. Let us look at this
in more detail. A typical Dirac spectrum is shown in Fig. \ref{fig1}.
The average number of eigenvalues in the rectangular contour
in this figure is $\rho(\lambda) l$. If
we integrate the resolvent along this contour  we find
\be
\oint G(z) = il(G(i\lambda+\epsilon) - G(i\lambda -\epsilon))
=2 \pi i \rho(\lambda) l,
\ee
where the second identity follows from Cauchy's theorem.
Using the symmetry of the spectrum we obtain for $\epsilon \to 0$
\be
{\rm Re } G(i\lambda+\epsilon) = \pi \rho(\lambda).
\ee
Near the center of the spectrum the imaginary part of the resolvent
is negligible. Using that
the chiral condensate is related to  the resolvent by
\be
\langle \bar \psi \psi \rangle
= - \lim_{m\to 0}\lim_{V\to \infty} \frac 1V G(m),
\ee
immediately results in the Banks-Casher relation (\ref{bankscasher}).
The order of the limits in (\ref{bankscasher}) is important. First we take the
thermodynamic limit, next the chiral limit (and, finally, the continuum
limit).

The resolvent of the QCD Dirac spectrum can be obtained from 
\be
G(z, z',m_f) =
 \left . \frac {\partial}{\partial z}
\right |_{z = z'} \log Z_\nu^{\rm pq}(z,z',m_f),
\ee
with the so called partially quenched  QCD partition function 
given by
\be
Z^{\rm pq}_\nu(z,z',m_f)  ~=~ 
\int\! dA
~\frac{\det(D + z)}{\det(D + z')}\prod_{f=1}^{N_{f}}
\det(D + m_f) ~e^{-S^{\rm YM}} ~.
\label{pqQCD}
\ee
For $z=z'$ this partition function  coincides with the
QCD partition function. 

In addition to the regular quarks, the partition function (\ref{pqQCD})
has  additional bosonic and fermionic ghost quarks. 
Our aim is to find the chiral Lagrangian corresponding to (\ref{pqQCD}).
If we are successful, we have succeeded in deriving a generating function
for the 
infrared limit of the QCD Dirac spectrum.

\subsection{Microscopic Spectral Density}

An important consequence of the Bank-Casher formula (\ref{bankscasher})
is that the eigenvalues near zero virtuality are spaced as
\be
\Delta \lambda = 1/{\rho(0)} = {\pi}/{\Sigma V}.
\label{spacing}
\ee
For the average position of the smallest nonzero eigenvalue 
we obtain the estimate
\be
\lambda_{\rm min} = {\pi}/{\Sigma V}.
\ee
This should be contrasted with the eigenvalue spectrum
of the non-interacting
Dirac operator. Then the eigenvalues are those of a free Dirac particle in a 
box with eigenvalue spacing equal to  $\Delta \lambda \sim 1/V^{1/4}$
for the eigenvalues near $\lambda = 0$.
Clearly, the presence of gauge fields leads to a strong modification of
the spectrum near zero virtuality. Strong interactions result in the
coupling of many degrees of freedom leading to extended states 
and correlated eigenvalues.
Because of asymptotic freedom, the spectral density of the Dirac operator for
large $\lambda$ behaves as $V\lambda^3$. In Fig. 2 we show a plot of
a typical average spectral density of the QCD Dirac operator for 
$\lambda \ge 0$. The spectral density for negative $\lambda$ is obtained 
by reflection with respect to the $y$-axis. More discussion of this figure
will be given in section \ref{parti5}.

\begin{figure}[ht] 
\centerline{\includegraphics[height=5cm]{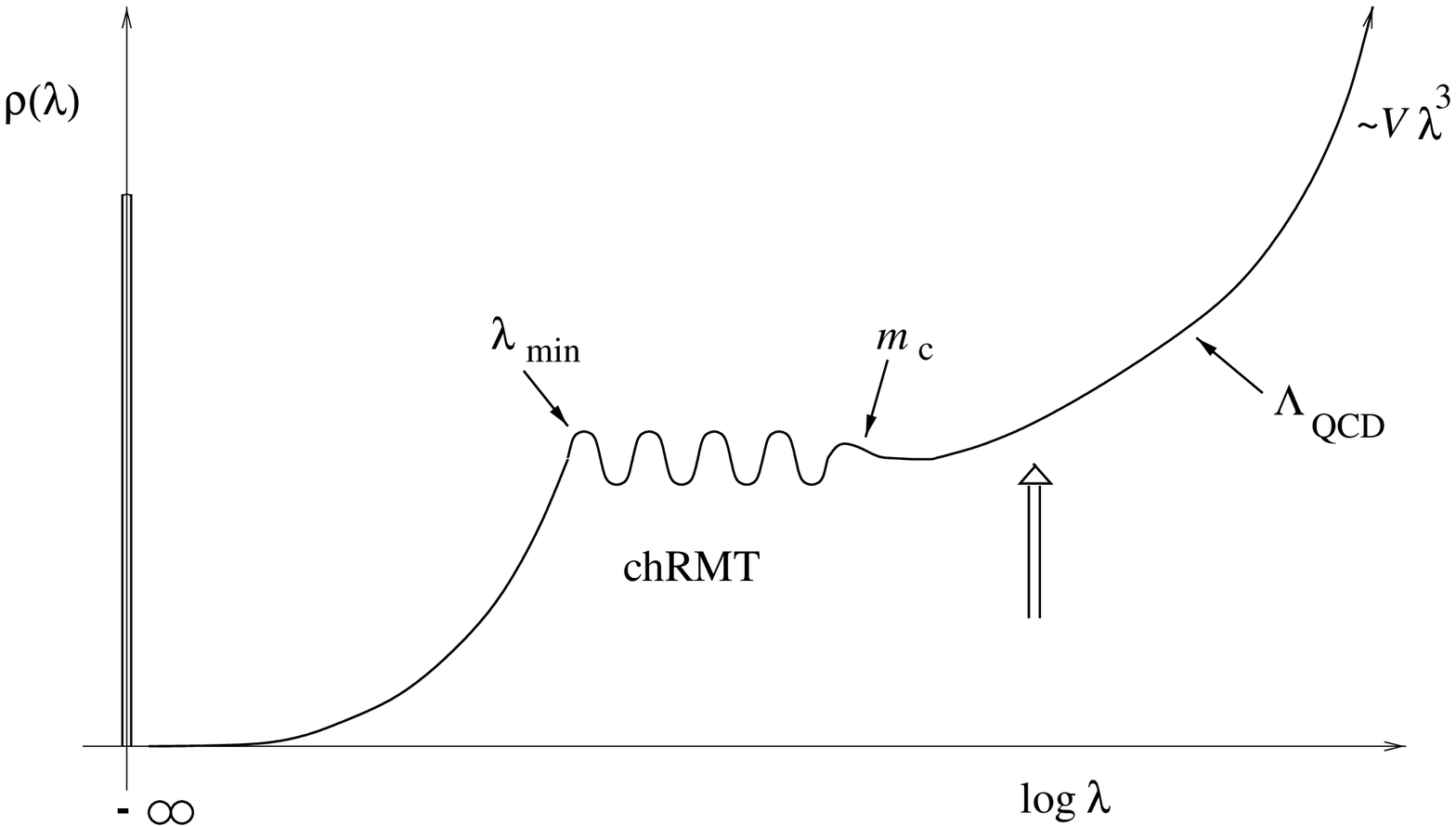}}   
\caption{}{\small Schematic picture of the average spectral density of 
QCD Dirac operator. (Taken from \cite{minn00}.)}

\label{fig1a}
\end{figure}

Because the eigenvalues near zero are spaced as $\sim 1/\Sigma V$
it is natural to introduce 
the microscopic spectral density \cite{SV}
\be
\rho_s(u) = \lim_{V\rightarrow \infty} \frac 1{V\Sigma} 
\rho(\frac u{V\Sigma})\quad {\rm with} \quad u = \lambda V \Sigma.
\label{rhosu}
\ee
We expect that this limit exists and converges to a universal function
which is determined by the global symmetries of the QCD Dirac operator.
In section \ref{chrmt},
we will calculate $\rho_s(u) $ both for the simplest theory in this
universality class, which is chiral Random Matrix Theory (chRMT), and 
for the partial quenched chiral Lagrangian which describes the low-energy
limit of the QCD partition function. We will 
find that the two results coincide
below the Thouless energy.

\section{Low Energy Limit of QCD}
\label{low}
In this section we derive the chiral Lagrangian that  provides an exact 
description of QCD at low energies. 

    \subsection{The chiral Lagrangian}
\label{lowen1}

For light quarks the low energy limit of QCD is well understood. 
Because of confinement, the only light degrees of freedom are
the pseudo-scalar mesons, that interact
according to a chiral
Lagrangian. To lowest order in the quark masses and the momenta, 
the chiral Lagrangian is completely determined by chiral symmetry and 
Lorentz invariance. In the
case of $N_f$ light quarks with chiral symmetry breaking according to 
$SU_L(N_f) \times SU_R(N_f) \rightarrow SU_V(N_f)$ the Goldstone 
fields (pseudo-scalar mesons)
are given by $U \in SU(N_f)$. Under an  $SU_L(N_f) \times SU_R(N_f)$
transformation of the quark fields given in (\ref{flavorsym}), 
the Goldstone fields  $U$ transform in the same way as
the chiral condensate
\be
U \to U_R U U_L^{-1} .
\ee
The symmetry (\ref{flavorsym}) is broken the mass term. However, the full 
symmetry can be restored if we also transform the mass term as
\be
M_{RL} \to U_R M_{RL} U_L^{-1},\qquad 
M_{LR} \to U_L M_{LR} U_R^{-1} .
\label{umass}
\ee
The low energy effective theory should have the same invariance 
properties. To leading order it contains terms
of second order in the momenta and of first order in the
quark mass matrix. The invariant terms are:
\be
{\rm Tr} (\partial_\mu U \; \partial_\mu
U^\dagger),\qquad  {\rm Tr} (M_{RL}  U^\dagger), \qquad 
{\rm Tr} (M_{LR} U).
\ee
Since the QCD partition function is invariant under $M_{RL} 
\leftrightarrow M_{LR}$, the effective partition function should
also have this symmetry. The action of the Goldstone fields is
therefore given by the so called 
Weinberg Lagrangian \cite{Weinberg,GL}
\be
{\cal L}_{\rm
eff}(U)=\frac{F^2}{4}  \;{\rm Tr} (\partial_\mu U  \partial_\mu
U^\dagger)-\frac{\Sigma}{2} \; {\rm Tr} (M_{RL} U^\dagger+
M_{LR} U),
\ee
where $F$ is the pion decay constant, and $\Sigma$ is the chiral condensate.
The Goldstone fields can be parametrized as
$
 U={\rm exp} (i\sqrt 2 \Pi_a t^a/F),
$
with the generators of $SU(N_f)$ normalized as  ${\rm Tr}\, t^a t^b
=\delta^{ab}$.
This chiral  Lagrangian has been used extensively
for  the analysis of pion-pion scattering amplitudes \cite{GL}.

To lowest order in the pion fields we find for equal quark masses $m$
\be
{\cal L}_{\rm eff}(U) = \frac 12 \partial_\mu \Pi^a \partial^\mu \Pi^a
              -N_f \Sigma m
+\frac{\Sigma m}{F^2} \Pi^a \Pi^a .
\ee
This results in the pion propagator $1/(p^2 +m_\pi^2)$ with pion mass
given by the Gellmann-Oakes-Renner relation
\be
m_\pi^2 =\frac{2m\Sigma}{F^2}.
\ee
It also illustrates the identification of $\Sigma$ as the chiral
condensate.

\subsection{The Low Energy Limit of $Z^{\rm pq}_\nu$}
\label{parti4}

The low-energy limit of the partially quenched QCD partition function
can be derived along the same lines as the derivation of the chiral
Lagrangian obtained in previous section. In this case,
ignoring convergence questions for the moment, the global flavor symmetry
of (\ref{pqQCD}) is given by by the supergroup
\be
Gl_R(N_f+1|1) \times Gl_L(N_f+1|1).
\label{gl}
\ee
This reflects that we have $N_f+1$ fermionic quarks and 1 bosonic quark.
We already have seen that convergence requirements restrict 
the axial symmetry for bosonic quarks
to $Gl(1)/U(1)$. Although the axial 
flavor symmetry
group of the fermionic quarks is not a priori determined by convergence
requirements
we will see in this section that supersymmetry necessarily imposes 
that this symmetry group is compact, i.e. equal to $U(N_f)$.

Under transformation (\ref{gl})
the quarks fields with $N_f+1$ fermionic components and one bosonic
component, transform as ($U_R, \, U_L \in Gl(N_f+1|1)$)
\be
\psi_R \rightarrow U_R\psi_R, \quad \psi_L \rightarrow U_L\psi_L,\quad
\bar\psi_R \rightarrow \bar\psi_RU_R^{-1}, \quad \bar\psi_L 
\rightarrow \bar \psi_L U_L^{-1}.
\ee
The subscripts refer to the
right-handed (R) or left-handed (L) quarks. 
For $M= 0$ and $\nu =0$ this is a symmetry of the QCD action.
For $M\ne 0$ this symmetry can be restored if we also transform
the mass term according to
\be
M_{RL} \to U_R M_{RL} U_L^{-1},\qquad
M_{LR} \to U_L M_{LR} U_R^{-1}.
\ee
In the sector of topological charge $\nu$
the partially quenched partition function transforms as
\be
Z_\nu^{\rm pq}(M_{RL},M_{LR}) 
\rightarrow {\rm Sdet}^\nu [U_R U_L]^{-1} Z_\nu^{\rm pq}
(M_{RL},M_{LR} ).
\label{pqtrans}
\ee

The Goldstone bosons corresponding to the breaking 
of the axial subgroup $Gl_A(N_f+1|1)$
transform as
$
Q\to U_R Q U_L^{-1}.
$
If we factorize the Goldstone fields into the zero momentum modes
$Q_0$ and the nonzero momentum modes $Q(x)$ as
\be
Q = Q_0 Q(x),
\ee
one can easily show that
the low energy effective
partition function with the above
transformation properties is given by

\be
Z_\nu^{\rm pq}(M) = \int_{Q \in Gl(N_f+1|1)} dQ {\rm Sdet}^\nu (Q_0) 
e^{-\int d^4x {\cal L}^{\rm pq}(Q)},
\ee 
where (see \cite{EfetovBook,mex} for definitions of the superdeterminant 
Sdet and and the supertrace Str)
\be
{\cal L}^{\rm pq}(Q) = 
\frac {F^2}4{\rm Str } \del_\mu Q^{-1} \del_\mu Q - 
\frac \Sigma 2 {\rm Str}(M_{RL} Q^{-1}) 
- \frac \Sigma 2 {\rm Str} (M_{LR} Q).\,\,
\label{chpq}
\ee
We already have seen that the boson-boson block of $Gl(N_f+1|1)$ is
 $Gl(1)/U(1)$. If we parameterize the field $Q$ as
\be
Q= e^{\sum_k T_k \pi_k/F},
\ee
with $T_k$ the generators of $G(N_f+1|1)$, to second order in the
Goldstone fields  the mass term is given by
$
{\rm Str} (\Sigma M \sum_k T_k^2 \pi_k^2/F^2).
$
Let us take $M$ diagonal positive definite. 
Because of the supertrace there is 
a relative minus sign between the boson-boson and 
fermion-fermion modes. The boson-boson modes are noncompact
and require that the overall minus sign of the mass term is negative. 
In order to avoid tachyonic fermion-fermion Goldstone modes, we have
to compensate the minus sign of the supertrace. This can be done by
choosing the parameters that multiply  the fermion-fermion generators
purely imaginary. This corresponds to 
a compact parametrization of the fermion-fermion Goldstone
manifold. This integration manifold is the maximum Riemannian
submanifold \cite{class} of $Gl(N_f+1|1)$ and will be denoted by
$\hat {Gl}(N_f+1|1)$.

\subsection{The Mesoscopic Limit of QCD} 
\label{parti5}

 In chiral perturbation theory, the different domains of validity where
analyzed           by Gasser and Leutwyler \cite{GaL}. A similar analysis
applies to partially quenched chiral perturbation theory \cite{Vplb}. 
The idea is
as follows. The $Q$ field can be decomposed as \cite{GaL}
\be
Q = Q_0 e^{i\psi(x)},
\ee
where $Q_0$ is a constant (zero-momentum) field. 
For momenta $p=\pi k/ L$ with $k$ integer, the kinetic term of
the $\psi$ fields behaves as 
\be
\frac 12  \partial_\mu \psi^a(x) \partial_\mu \psi^a(x)  \sim L^{-2} 
\psi^a(x)\psi^a(x).
\ee
We observe that the magnitude of the fluctuations of the $\psi^a(x)$ 
fields are of order  $1/L$ which justifies a perturbative expansion of
$\exp(i \psi(x))$. The fluctuations of the zero momentum modes
are only limited by the mass term 
\be
\frac 12  V \Sigma {\rm Str} {M}(Q_0+ Q_0^{-1}).
\ee
For quark masses 
$m \gg 1/V\Sigma$, the field $Q_0$ fluctuates close to the identity and
the $Q_0$ field can be expanded around the identity as well. If 
$m \ll \Lambda_{\rm QCD}$ we are in the  domain of chiral perturbation theory.
For
\be
\frac {\Sigma m}{F^2} \ll\frac 1{\sqrt V}
\ee
the fluctuations of the zero momentum modes dominate the fluctuations of 
the nonzero momentum modes, and only the contribution from the 
zero momentum modes 
has to be taken into account for the calculation of an observable.
In this limit the so called finite volume partition function is given by
\cite{GaL,LS}
\be
Z^{\rm eff}_{N_f}({M},\theta)
\sim \int_{U\in SU(N_f)} dU e^{V\Sigma {\rm Re\,Tr}\,
{ M} U^\dagger e^{i\theta/N_f}},
\label{zeff}
\ee
where the $\theta$-dependence follows from the dependence of the QCD
partition function on the combination  $m e^{i\theta/N_f}$ only
(see section \ref{symme2}). 
We emphasize that any theory with the same pattern of chiral symmetry
breaking as QCD can be reduced to  the same extreme infrared limit.

The effective partition function at fixed $\nu$ follows by Fourier
inversion \cite{LS}
\be
Z_{\nu,N_f}^{\rm eff}(M) = \frac 1{2\pi} \int_0^{2\pi} d \theta 
e^{-i\nu \theta} Z^{\rm eff}(M,\theta).
\ee
Combining the integral over $SU(N_f) $ and the integral over $U(1)$ 
we find that
\be
Z^{\rm eff}_{\nu,N_f}(M) = \int_{U(N_f)} {\det}^\nu (U) 
e^{V\Sigma {\rm Re\,Tr}\,
{M} U^\dagger }.
\label{zeffnu}
\ee

The same arguments apply to the partially quenched chiral Lagrangian.
There is an important difference. The mass of the ghost-quarks 
is an external parameter which can take on any value we wish.
Indeed, the mass of the Goldstone modes containing
these quarks is given by
\be
M_{zz}^2 = \frac {2 z \Sigma}{F^2}.
\ee 
Therefore, independent of the quark masses there is always a domain
where the fluctuations of the zero momentum modes dominate the 
fluctuations of the nonzero momentum modes. This domain is given by
\cite{Vplb}
\be
 z \ll \frac{F^2}{\Sigma L^2}\equiv m_c.
\label{domain}
\ee
In this domain, the Compton wavelength of the Goldstone bosons with mass
$M_{zz}$ is much larger than the size of the box.
Because the time scale conjugate to $m_c$ is of the order of the diffusion
time across the length of the box, this domain
is known as the ergodic domain. 
For the non-Goldstone modes
not to contribute to the partition function we have to require that
$L \gg 1/ \Lambda_{\rm QCD}$.

In the Dirac spectrum we can thus distinguish three 
important scales: 
\be
\lambda_{\rm min} \ll m_c \ll \Lambda_{\rm QCD}
\label{dompq}
\ee
(see Fig. \ref{fig1a}).
For $z \ll m_c$ we are in the zero momentum sector of the theory. If $z$
is of the order of $\lambda_{\rm min}$ or less
we have to take into account quantum fluctuations to all orders. For 
$\lambda_{\rm min} \ll z \ll m_c$, the integral over zero modes can
be calculated perturbatively by a loop expansion.
For $m_c \ll z \ll \Lambda_{QCD}$, chiral perturbation
theory still applies, but the zero momentum modes no longer dominate the
partition function. 
For $z \gg \Lambda_{QCD}$, 
chiral perturbation theory is not applicable to the spectrum of
the Dirac operator. 

In the ergodic domain the QCD partition
function in the sector of topological charge $\nu$ is given by \cite{OTV}
\be
Z_\nu^{\rm pq}( M) =
\int_{Q\in \hat{Gl}(N_f+1|1)} dQ \,{\rm Sdet}^\nu Q \, e^{
 V\frac{\Sigma}{2} \; {\rm Str} (M Q+M Q^{-1})}.
\label{superpart}
\ee
The number of QCD Dirac eigenvalues that is described
by this partition function is of the order 
$m_c/\Delta \lambda= F^2L^2$.
This  number increases linearly in $N_c$ for $N_c\to \infty$ which
was recently found in lattice simulations \cite{neuberger}.

In section (\ref{parti7}) we will study this partition function in the 
quenched limit ($N_f = 0$) and show it coincides with the chRMT 
result \cite{OTV,DOTV}.

\subsubsection{Comparison to Disordered Systems}

In the book by Efetov \cite{EfetovBook} it is shown that the diffusion of
electrons in a disordered medium can be described  by the effective action
\be
F(Q) = 
\int d^d x [ 
\frac {\pi \nu} 8 D {\rm Tr} (\nabla Q)^2 
-\frac {\pi i\nu \omega}4 {\rm Tr} \Lambda Q],
\label{chdis}
\ee
where $Q$ are the Goldstone fields, $\nu$ is the density of states,
$D$ is the diffusion constant and $\omega $ is the energy difference
between the advanced and the retarded Green's functions. 
The matrix $\Lambda$ is a diagonal matrix
with matrix elements $\pm 1$ corresponding to the causal character of the
Green's functions. 
The Goldstone
bosons arise because of the spontaneous breaking of the symmetry between
the advanced and retarded Green's functions. 

If we compare this effective action to the chiral Lagrangian
(\ref{chpq}) we can make the identification
\be
\frac {F^2}4 \leftrightarrow \frac{\pi \nu D}8 , 
\qquad \frac{\pi \omega \nu}4 \leftrightarrow 
\frac {M\Sigma}2, \qquad
\nu \leftrightarrow \frac{\rho(E)}V,
\ee
which can be rewritten as
\be
M \leftrightarrow \frac \omega 2, \qquad \Sigma \leftrightarrow \pi\nu, 
\qquad F^2 \leftrightarrow \frac {\pi\nu D}2.
\ee
The domain where the kinetic term factorizes from the partition function
is therefore given by \cite{Vplb,jprl}
\be
L^2\ll \frac {F^2}{M\Sigma} \leftrightarrow\frac D\omega. 
\ee
In the theory of disordered mesoscopic systems the corresponding 
energy  scale is
known as the Thouless energy. It is 
defined by \cite{Altshuler,Montambaux}:
\be
E_c = \frac{\hbar D}{L^2},
\ee
The time conjugate to $E_c$ is the time scale over
which an electron diffuses across the sample. Therefore, the domain 
where $\hbar \omega \ll E_c$ is known as the ergodic domain.
The
time scale in mesoscopic physics corresponding to
$\Lambda_{QCD}$ is the elastic scattering time $\tau_e$. The domain in between
$E_c$ and $\hbar/\tau_e$ 
is known as the diffusive domain. This domain is
characterized by diffusive motion of electrons in the disordered sample
described by the Lagrangian (\ref{chdis}).

\section{Chiral RMT and the QCD Dirac Spectrum}
\label{chrmt}
\subsection{ The chiral ensembles}
\label{rando2}

The chiral ensembles
are defined as the ensembles of $N\times N$
Hermitian matrices with block structure \cite{SV,V}
\be
\label{hc}
D= \left ( \begin{array}{cc} 0 & iC \\ iC^\dagger & 0 \end{array} \right ),
\label{chdir}
\ee
and probability distribution given by (for equal quark masses $m$)
\be 
P(C)dC = {\cal N}{\det}^{N_f}(D+m)
e^{-\frac {N\beta_D}{4} {\rm Tr} C^\dagger C}
dC.
\label{chRMTprob}
\ee
The integration measure $dC$ is the product of differentials of the independent
parts of the matrix elements of $C$, and $N_f$ is a real parameter
(corresponding to the number of quark flavors in QCD).
The matrix $C$ is a rectangular $n \times (n+\nu)$ matrix and $N=2n+\nu$.
The nonzero eigenvalues of the matrix $D$ occur in pairs $\pm \lambda_k$.
This can be seen as follows. If 
\be
D\left ( \begin{array}{c} a \\ b \end{array} \right )
= \lambda \left ( \begin{array}{c} a \\ b \end{array} \right )
\quad
{\rm then} \quad
D\left ( \begin{array}{c} a \\ -b \end{array} \right )
= -\lambda \left ( \begin{array}{c} a \\ -b \end{array} \right ).
\ee
Generically, the matrix $D$ in (\ref{chdir})
 has exactly $|\nu|$ zero eigenvalues.
For this reason, $\nu$ is identified as the topological quantum
number. The normalization constant of the probability distribution is
denoted by ${\cal N}$.  We can distinguish 
ensembles with real, complex, or quaternion real matrix elements. They
are denoted by $\beta_D= 1$, $\beta_D = 2$, and $\beta_D = 4$, respectively.
In addition to the global symmetries of QCD, this partition function 
has the large unitary invariance 
\be
C \rightarrow U C V^{-1},
\label{uninv}
\ee
where $U$ and $V$ are orthogonal, unitary, or symplectic matrices,
respectively. Therefore, the corresponding ensembles are known
as the  chiral Gaussian Orthogonal Ensemble (chGOE), the    
  chiral Gaussian Unitary Ensemble (chGUE), and the  chiral
  Gaussian Symplectic Ensemble (chGSE), in this 
order.

Using the invariance (\ref{uninv}) it is always possible to decompose $C$ as
\be
C = U \Lambda V^{-1},
\label{decomc}
\ee
where $\Lambda$ is a diagonal matrix  with $\lambda_{k}\ge 0$.
The joint probability distribution for the eigenvalues
is obtained by transforming to $\Lambda$, $U$ and $V$ 
as new integration variables. 
 The Jacobian is given by
\be 
J \sim \prod_k \lambda_k^{\nu\beta_D  -1}
\prod_{k<l} |(\lambda^2_k -\lambda^2_l)|^{\beta_D} 
\label{jchiral}
\ee
resulting in the joint eigenvalue distribution
\be
P(\{\lambda\}) d \{\lambda\} = {\cal N} |\Delta(\{\lambda^2\})|^{\beta_D}
\prod_k \lambda_k^\alpha(\lambda_k^2+m^2)^{N_f} 
e^{-N\beta_D \lambda_k^2/4} d\lambda_k,
\label{joint}
\ee
where  $\alpha= \beta_D -1 +\beta_D \nu $.
We note that the distribution of the eigenvectors factorizes
from the distribution of the eigenvalues.

\subsection{Mathematical Methods}

In this section we will discuss
the orthogonal polynomial method, the resolvent expansion method,
the replica trick and the supersymmetric method. These
methods are widely used in Random Matrix Theory.  

\subsubsection{Resolvent Expansion Methods}

These methods are based on expanding the resolvent in a geometric series
\be
G(z) = \langle{\rm Tr}  \frac 1{z-H}\rangle  = N\frac 1z + 
\langle {\rm Tr} \frac 1z H \frac 1z
\rangle +\langle {\rm Tr}\frac 1z H \frac 1z H \frac 1z\rangle + \cdots \,.
\ee
In the large $N$ limit the averages are given by a sum of planar diagrams.
Let us illustrate this for the GUE. In this case the ``propagator'' is
given by
\be
\langle H_{ij} H_{kl}\rangle = \frac 1N \delta_{il} \delta_{jk}.
\ee
For example, as was explained in the course of Di Francesco \cite{francesco}, 
for  ${\rm Tr} H^4$ term we have two planar diagrams of 
order $N^3$ and  one diagram of order $N^2$.

\subsubsection{The Orthogonal Polynomial Method}
The oldest method
is the orthogonal polynomial method \cite{Mehta}. In principle, one obtains
expressions that are exact for finite size matrices. The drawback
of this method is that it requires a probability distribution that
is invariant under basis change of the random matrix. In general
the probability density can be written as
\be
P(x_1, \cdots,x_n) = \Delta^{\beta_D}(\{x_k\}) \prod_{k=1}^n w(x_k),
\ee
where $w(x)$ is a weight function and the Vandermonde determinant is given by
\be
\Delta(\{x_k\}) = \prod_{k>l} (x_k -x_l).
\ee

The method is based on the identity
\be
\Delta(\{x_k\})=\left |
\begin{array}{ccc} 1 & \cdots & 1 \\
x_1 & \cdots & x_n\\
\vdots && \vdots\\
x_1^{n-1}& \cdots & x_n^{n-1}
\end{array}\right | 
=
\left| \begin{array}{ccc} P_0(x_1) & \cdots & P_0(x_n) \\
P_1(x_1) & \cdots & P_1(x_n)\\
\vdots & & \vdots\\
P_{n-1}(x_1) &\cdots &P_{n-1}(x_n) 
\end{array}\right | ,
\ee
where the $P_k$ are monic orthogonal polynomials defined by
\be
\int dx w(x)P_k(x)P_l(x) = h_k \delta_{kl}.
\ee
Because of these relations, integrals over the eigenvalues can be performed
by means of orthogonality relations. In the next section we
illustrate this method by the calculation of the microscopic spectral
density for the chGUE.

\subsubsection{The Replica Trick}

The replica trick is based on the identity
\be
G(z) &=&\frac 1V \langle {\rm Tr} \frac 1{z+iD}\rangle
 = \lim_{r\to 0} \frac 1{Vr} \del_z \langle {\det}^r(iD+z) \rangle.
\label{repdef}
\ee
The recipe is to calculate the partition function for positive 
or negative integer values of $r$ and then analytically continue to
$r =0$. For positive (negative) integer values of $r$ the average
determinant can be calculated by rewriting it as 
a Grassmann (complex) Gaussian integral. Then $D$ appears linear
in the exponent which allows us to perform the average for a
Gaussian distribution of $D$.
The replica trick works without problems for perturbative calculations
\cite{dam-split-rep}
but usually fails in nonperturbative calculations \cite{critique}.
As example consider the following expression for the the
modified Bessel function $I_\nu(z)$:
\be
I_\nu(z) = \frac 1\pi \int_0^\pi e^{z\cos\theta} \cos\nu\theta d\theta
-\frac {\sin\nu \pi}{\pi} \int_0^\infty e^{-z\cosh t -\nu t}dt.
\ee
We would have missed the second term if we calculate the Bessel function only
for integer values of $\nu$. The replica trick can be made to work
if we consider a family of partition functions related by a Toda
lattice equation. This will be discussed in detail in the next two
lectures.

\subsubsection{The Supersymmetric Method}

The supersymmetric method \cite{Efetov} is based on the identity
\be
G(z) = \frac 1V \del_z
\left . \left\langle \frac{\det(iD+z)}{\det(iD+z')}\right \rangle
\right |_{z'=z}.
\ee
The determinant can be written as a Grassmann integral and the inverse
determinant as a complex integral. For $z'=z$ this partition
function has an exact supersymmetry. The advantage of this method is that
it is mathematically rigorous, but it requires a deep understanding
of super mathematics. For example, finite expressions can be obtained
from singular terms that do not depend on the Grassmann variables
(and are zero upon integration). For a discussion of these
so-called Efetov-Wegner terms we refer to the literature
\cite{Zirnhal,mex}.

\subsection{The Microscopic Spectral Density of the chGUE}

In this subsection we calculate the microscopic spectral density
by the orthogonal polynomial method
\cite{VZ} and  the supersymmetric method \cite{OTV,DOTV}.

\subsubsection{Orthogonal Polynomials}

For the chGUE the joint probability distribution
only depends on the square of the eigenvalues. In terms of
$x_k = \lambda_k^2$ the weight function for $m_f =0$ is  given by
\be
w(x) = (x n\Sigma^2)^a e^{-n\Sigma^2 x},
\ee
with
$
a = N_f + |\nu|.
$
The monic orthogonal polynomials corresponding to this weight function can
be expressed in terms Laguerre polynomials
\be
P_k(x) = \frac {(-1)^k k!}{(\Sigma^2 n)^k} L_k^a(x\Sigma^2 n)
\ee
with normalization constants $h_k$ given by
$h_k = {k!(k+a)!}/{(n\Sigma^2)^{2k+1}}$.
The eigenvalue density is given by (with $c$ a constant)
\be
\rho(x_1) &=& c\int \prod_{k=2}^n [w(x_k) dx_k]
\left | \begin{array}{ccc} P_0(x_1) & \cdots & P_0(x_n) \\
\vdots && \vdots\\
P_{n-1}(x_1) &\cdots& P_{n-1}(x_n)
\end{array}\right |^2, \nn \\
&=& \sum_{\sigma \pi} {\rm sg}(\sigma\pi)\prod_{k=2}^n [w(x_k) dx_k]
P_{\sigma(0)}(x_1) \cdots P_{\sigma(n-1)}(x_ n) \nn \\
&&\times
P_{\pi(0)}(x_1) \cdots P_{\pi(n-1)}(x_ n),\nn \\
&=&(n-1)! \prod_{l=0}^{n-1} h_l
\sum_{k=0}^{n-1}\frac 1{h_k} P_k^2(x_1) w(x_1).
\ee
The microscopic spectral density is obtained by taking the limit 
$n \to \infty$ for fixed $z = 2n\Sigma \lambda=2n\Sigma \sqrt x$.
In this limit the weight function is given by $w(x) = (x\Sigma^2n)^a$,
and the Laguerre polynomials behave as
\be
L_k^a(x\Sigma^2n) \to k^a (x\Sigma^2nk)^{-a/2}J_a(2\Sigma\sqrt{xnk}).
\ee
In the limit $n \to \infty$,
the sum can be replaced by an integral resulting in 
the microscopic spectral density \cite{VZ}
\be
\rho(z) &\sim& z\sum_{k=0}^{n-1} J_a^2\left (z\sqrt{\frac kn}\right )
           \approx z\int_0^1 tdt J_a^2(zt),\nn\\
           &=& 2z(J_a^2(z) -J_{a+1}(z)J_{a-1}(z)).
\label{microop}
\ee

\subsubsection{Supersymmetric Method}

\label{parti7}

In this section we evaluate the resolvent of QCD for
the simplest case of $N_f = 0$ and $\nu =0$ in the
domain $z\ll F^2/\Sigma L^2$. In this domain
the partition function is given by (see (\ref{superpart}))
\be
Z(J) = \int_{Q\in \hat{Gl}(1|1)} dU \exp\left [ \frac {\Sigma V}2 {\rm Str}
\left ( \begin{array}{cc} z +J &0 \\ 0 & z \end{array} \right )
(Q + Q^{-1})\right ],
\ee
where the integration is over the maximum super-Riemannian sub-manifold
of $Gl(1|1)$. This manifold is parametrized by
\be
Q =\exp {\left ( \begin{array}{cc} 0 &\alpha \\ \beta 
& 0 \end{array} \right )}
 \left ( \begin{array}{cc} e^{i\phi} &0 \\ 0 & e^s \end{array} \right ).
\ee
The integration measure is the Haar measure which in 
terms of this parameterization
is given by (with $\delta Q \equiv Q^{-1} dQ$)

\be
{\rm Sdet } \frac {\delta Q_{kl}}{
\delta \phi \, \delta s\, \delta \alpha \, \delta \beta} 
\, \,d\alpha d\beta d\phi ds.
\ee

It is straightforward to calculate the Berezinian going from the
variables $\{\delta Q_{11},\,\delta Q_{22},\,
\delta Q_{12},\,\delta Q_{21}\}$ to the variables 
$\{\delta \phi, \, \delta s,\, \delta \alpha, \, \delta \beta\}$. The 
derivative matrix is given by
\be
B=\frac {\delta Q_{kl}}{
\delta \phi \, \delta s\, \delta \alpha \, \delta \beta} =
\left (\begin{array}{cccc} 
i & 0 & \frac \beta 2&
\frac \alpha 2\\
0  &1 & \frac \beta 2 &\frac \alpha 2\\
0 & 0& e^{s-i\phi}& 0 \\
0 & 0 & 0 &e^{-s+i\phi} 
\end{array}\right).
\ee
Using the definition of the graded determinant one simply finds that
${\rm Sdet} B = i$. Up to a constant, the integration measure is thus given by
$d\phi ds d\alpha d\beta$. 

We also need
\be
\frac 12(Q +Q^{-1} ) = \left ( \begin{array}{cc} 
\cos \phi (1 +\frac {\alpha \beta} 2 ) & \alpha (e^s - e^{-i\phi})\\
\beta(e^{i\phi} - e^{-s} ) & \cosh s ( 1- \frac{\alpha \beta }2) 
\end{array} \right ).
\ee
After differentiating with respect to the source term  
$(G(z) = \partial_J\log Z(J) |_{J=0}) $ this results in
(with $x = V\Sigma z$)
\be
\frac{G (z)}{V\Sigma} =  \int \frac{d\phi ds d\alpha d\beta}{2\pi}
\cos \phi (1 +\frac {\alpha \beta} 2 ) 
e^{x \cos \phi (1 +\frac {\alpha \beta} 2 ) -
x\cosh s ( 1- \frac{\alpha \beta }2)}.\nonumber \\
\ee
With the Grassmann integral given by the coefficient of $\alpha\beta$ 
we  obtain
\be
\frac{G(z)}{V\Sigma} &=& \int \frac{ds d\phi}{4\pi} 
[\cos\phi  \nonumber +x(\cos\phi 
+ \cosh s)\cos\phi ]e^{x(\cos\phi - \cosh s )} .
\ee
All integrals can be expressed in terms of modified Bessel functions.
We find 
\be
\frac{G(z)}{V\Sigma} &=& I_1(x)K_0(x)
+\frac x2 ( I_2(x)K_0(x)+I_0(x)K_0(x)+ 2I_1(x)K_1(x)),\nn\\
\ee
which can be further simplified by the recursion relation
$
I_2(x) = I_0(x) -2 I_1(x)/x.
$
As final result we obtain \cite{Vplb,OTV,DOTV}
\be
\frac{G(z)}{V\Sigma} = x(I_0(x)K_0(x)+ I_1(x)K_1(x)).
\label{resnf0}
\ee

This calculation can be generalized to arbitrary $N_f$ and arbitrary $\nu$. 
The calculation for arbitrary $N_f$ with $m_f =0$ 
is much more complicated, but with a
natural generalization of the factorized parameterization, and using some known
integrals over the unitary group, 
one arrives at the following
expression in terms of modified Bessel functions
\be
\frac {G(z)}{V\Sigma} =\frac \nu x + x(I_{a}(x)K_{a}(x)
+I_{a+1}(x)K_{a-1}(x)),
\label{val}
\ee
where $a = N_f+|\nu|$. This result is in 
complete agreement with the resolvent obtained \cite{Vplb}
from integrating the microscopic spectral density  (\ref{microop}).

\begin{figure}[!h] 
\centerline{\includegraphics[height=10cm]{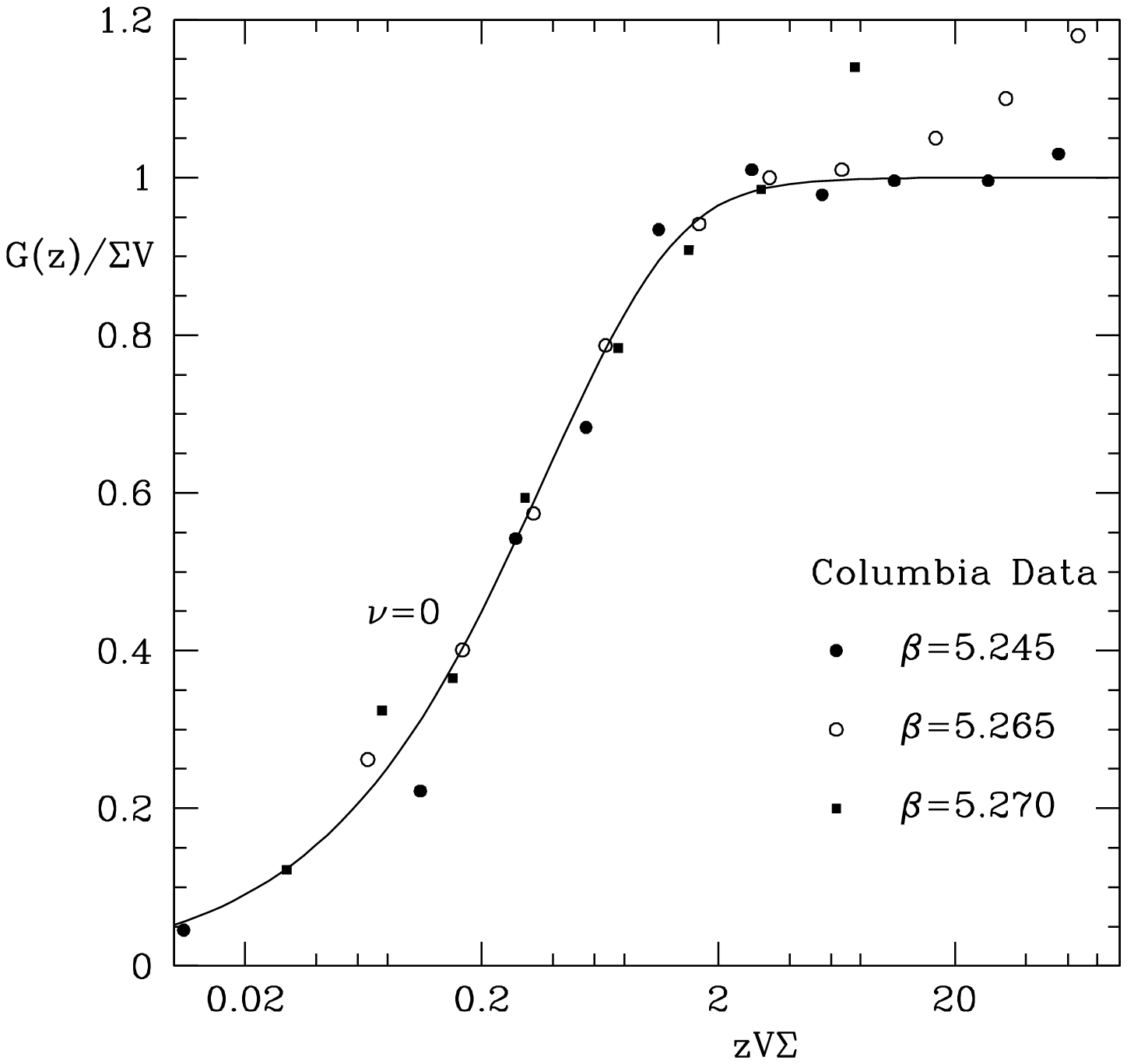}}   
\caption{}{\small The resolvent of quenched QCD. The points represent lattice
data obtained by the Columbia group, and the theoretical prediction
(\ref{resnf0}) is given by the solid curve.
(Taken from ref. \cite{minn04}.)}
\label{fig3}
\end{figure}

For $a=0$ this result is plotted in Fig. \ref{fig3}. We observe that,
below some scale, lattice QCD data obtained by the Columbia group
\cite{Christ}
closely follow this curve.
The predictions of chRMT or of the 
partially quenched chiral Lagrangian
have been studied by numerous lattice 
simulations \cite{anrev,poul,berbenni,neuberger}.
In all cases,
agreement has been found in the expected domain of applicability.

\section{Integrability and the QCD Partition Function}
\label{integ}
\subsection{Virasoro Constraints}
In this section we derive the small mass expansion of the QCD partition
function by means of recursion relations for the partition function
known as Virasoro constraints. The starting point is the QCD partition 
function in the ergodic regime given in (\ref{zeffnu}).
The quantities
\be
{\cal G}_\nu(t_k) = {\det}^{-\nu}(M) Z_{\nu,N_f}(M)
\ee
are invariant under the $U(N_f) \times U(N_f)$ 
transformations $M \to V_1 M V_2^{-1}$. (Here and below we drop
the superscript eff.)
Therefore, ${\cal G}_\nu(t_k)$ 
only depends on the
eigenvalues of $M^\dagger M$ which can be parameterized in terms of
the moments
\be
t_k \equiv \frac 1k {\rm Tr} \left( \frac{MM^\dagger}4 \right )^k .
\ee

A differential equation for the ${\cal G}_\nu(t_k)$ is obtained from the 
unitarity relation
\be
\frac 1{Z_{\nu,N_f}(M)}\sum_{a=1}^{N_f}\frac 
{\del^2 Z_{\nu,N_f}(M)}{\del {M_{ba}}
\del {M_{ac}^\dagger}}
=\frac 14 
\sum_{a=1}^{N_f} \langle U^\dagger_{ab}U_{ca}\rangle =\frac 14 \delta_{bc}.
\ee
Notice that the factor $\Sigma V$ is included in $M$. This relation
can be rewritten as
\be 
 \left [ \frac {\del^2}{\del M_{ba} \del M^\dagger_{ac}}
+\nu M^{-1}_{ab}\frac {\del}{\del M_{ac}^\dagger}
\right ] {\cal G}_\nu(t_k) = \frac 14 {\cal G}_\nu(t_k) \delta_{bc}.
\label{geq}
\ee
Using the chain rule
and the assumption that the matrix elements of $ (MM^\dagger)^{s-1}$
are independent for different values of $s$, 
we obtain  \cite{mmm}
\be
[ {\cal L}_s - \delta_{s,1}] 
{\cal G}_\nu(t_k) =0 \quad s \ge 1.
\label{virasoro}
\ee
The Virasoro operators ${\cal L}_s$ defined by
\be
{\cal L}_s = \sum_{k=1}^{s-1} \frac \del{\del t_k}\frac \del{\del t_{s-k}}
+\sum_{k \ge 1 } t_k \frac\del{\del t_{s+k}} +(N_f +\nu)\frac\del{\del t_s}
, \quad s \ge 1
\ee
satisfy the Virasoro algebra
\be
[{\cal L}_r, {\cal L}_s] = (r-s){\cal L}_{r+s}.
\ee
Therefore, {\em all} Virasoro constraints with $s \ge 1$  are satisfied if 
${\cal G}_\nu$ satisfies
\be
[{\cal L}_1 -1] {\cal G}_\nu = 0 \qquad {\rm and } 
\qquad {\cal L}_2 {\cal G}_\nu = 0.
\ee
This justifies the
independence assumption above Eq. (\ref{virasoro}).

\subsubsection{Solution of the Virasoro Constraints}

We can expand $G_\nu$ as 
\be
G_\nu = 1 + a_1 t_1 + a_2 t_2 + a_{11} t_1^2 + \cdots\, .
\ee
From the first Virasoro constraint we obtain
\be
{\cal L}_1 G_\nu &=& a_2 t_1 + (N_f + \nu)(a_1 + 2 a_{11} t_1) + \cdots\, ,
\nn \\
               &=& 1  + a_1 t_1 + a_2 t_2 + a_{11} t_1^2 + \cdots\, .
\ee
By equating the coefficients of the $t_k$ we find
\be
a_1 = \frac 1{N_f+\nu},\qquad
a_2 +2(N_f+\nu)a_{11} = a_1.
\ee
From the second Virasoro constraint,
$
{\cal L}_2 G_\nu = a_{11} + (N_f+\nu) a_2 =0,
$
we obtain
$
a_2 = - {a_{11}}/(N_f+\nu).
$
Continuing this way we can obtain all coefficients in the 
expansion of the $G(t_k)$. This results in the small mass
expansion of the partition function \cite{dam-split}
\be
\frac{Z_\nu^{\rm QCD}(M)}{  {\det}^\nu M}= [ 1 + \frac{{\rm Tr} MM^\dagger}{4(N_f+\nu)}
+\frac 1{32} \frac{{\rm Tr}(MM^\dagger)^2}{(N_f+\nu)((N_f+\nu)^2 -1)} 
+ \cdots ].\hspace*{0.3cm}
\label{smallvir}
\ee
An extension of this expansion to all three Dyson classes can
be found in \cite{denvir}.

The small mass expansion can be used to obtain sum rules for the 
inverse eigenvalues of the Dirac operator \cite{LS,dam-split}.
The QCD partition function can be expanded as (the prime indicates
that $\lambda_k \ne 0$)
\be
Z_\nu^{\rm QCD}(M) &=& m^{\nu N_f}
\left \langle {\prod}'_k \lambda_k^{2N_f} ( 1+ \frac {m^2}{\lambda_k^2})^{N_f}) 
\right \rangle_\nu \\
&=& m^{\nu N_f} (\langle {\prod}'_k \lambda_k^{2N_f}\rangle_\nu
+ m^2 N_f \langle {\prod}'_k \lambda_k^{2N_f} {\sum}'_k \frac 1{\lambda_k^2}
\rangle_\nu +\cdots)\nn 
\ee
This results in the expansion
\be
\frac{Z_\nu^{\rm QCD}(M)}{\lim_{m\to 0} m^{-\nu N_f}Z_\nu^{\rm QCD}(M)} = 
1 + m^2 N_f \langle {{\sum}'}_k\frac 1{\lambda_k^2}\rangle_{\nu}^{{\rm QCD}}.
\ee
By equating this expansion to the expansion (\ref{smallvir}) for
equal masses given by
\be
1 +\frac{N_f(V\Sigma)^2}{4(N_f+\nu)} m^2,
\ee
we obtain a  Leutwyler-Smilga sum rule \cite{LS} for the inverse Dirac
eigenvalues
\be
\frac 1{V^2}{{\sum}'}_k\frac 1{\lambda_k^2} = \frac {\Sigma^2}{4(N_f+\nu)}.
\ee

\subsubsection{Flavor-Topology Duality}
 
If we construct an $(N_f + \nu)\times(N_f +\nu)$ matrix $\bar M$
with $\bar M_{ij} = M_{ij}$ for $ i,j \le N_f$ and $\bar M_{ij} = 0$
otherwise, we have that ${\rm Tr} (M M^\dagger)^k ={\rm Tr}
(\bar M \bar M^\dagger)^k$. 
Since the Virasoro constraints only depend on the combination $N_f + \nu$
we have 
\be
{\det}^{-\nu}M
  Z_{\nu, N_f}(M) = Z_{\nu=0,N_f+\nu}(\bar M).
\label{ft}
\ee
This relation is known at the flavor-topology 
duality \cite{cam97}.

\subsection{$\tau$-Function}

The unitary integral in the QCD partition function can actually
be evaluated analytically for an arbitrary number of flavors. 
We will show that it can be rewritten as a $\tau$-function.
The unitary
matrix integrals can then be evaluated by means of a
Harish-Chandra-Itzykson-Zuber type integral and the use of flavor-topology
duality.

\subsubsection{Itzykson-Zuber Integral}

We consider the integral
\be
I = \int dU dV e^{\frac 12 {\rm Tr} (U^\dagger RVS +S V^\dagger R U)},
\label{IZch}
\ee
where $U \in U(N_1)$ and $V\in U(N_2)/U^{N_2}(1)$ and the integral is
over the Haar measure of these groups. The matrices $R$ and $S$ are
arbitrary rectangular complex matrices. Without loss of generality,
they can be taken diagonal with $R_{kk} = r_k>0$ and 
$S_{kk} = s_k>0$
and all other matrix element equal to zero.
Using 
the diffusion equation method one can derive the result
\cite{guhr-wettig} 
\be
I = c \prod_k (r_ks_k)^\nu \frac {\det I_\nu(r_ks_l)}
{\Delta(\{s_k^2\}) \Delta(\{r_k^2\})}.
\label{IZ-ch}
\ee
This result first appeared in the Russian literature \cite{ber-kar} 
as a solution
of the Laplace equation. It was proved independently 
in \cite{guhr-wettig}.

\subsubsection{The QCD Partition Function is a $\tau$-Function}

In this subsection we show that the finite volume QCD partition function
is a $\tau$-function. 
Using the flavor-topology duality (\ref{ft}) 
with $\bar M_{kk} = x_k$  for $k\le N_f$ we can write
\be
  Z_{\nu, N_f}(M) ={\det}^{\nu}M 
\int_{U\in U(N_f+\nu)} dU e^{\frac 12
{\rm Tr}(\bar M U^\dagger + \bar M^\dagger U)}.
\ee
This integral is exactly
(\ref{IZch})  with $R$ equal to the  $N_f \times (N_f+\nu)$
matrix with $r_k = x_k$. A finite result is obtained if the 
$N_f$ diagonal matrix elements of $S$ are expanded as
$s_k = 1 + \delta s_k$. 
For $\delta s_k \to 0$ we obtain
\be
I_\nu(x_k s_l) &= &\sum_{j=1}^{N_f} \frac{x_k^{j-1}}{(j-1)!} 
I_\nu^{(j-1)}(x_k)
(\delta s_l)^{j-1}
\equiv\sum_{j=1}^{N_f} A_{kj} B_{jl}.
\ee
with (the upper index between brackets such as $(k)$ denotes the $k$'th
derivative)
\be 
A_{kj} = x_k^{j-1} I_\nu^{(j-1)}(x_k), \qquad B_{jl} = \frac{(\delta s_l)^{j-1}}
{(j-1)!}.
\ee
Up to a constant, the determinant of $B$ is given by
\be
\det B \sim \Delta(\{(1+\delta s_k)^2\}),
\ee
and cancels against the denominator in (\ref{IZ-ch}).
We finally obtain the result
\be
  Z_{\nu, N_f}(M) 
&=& \frac{\det [x_k^{j-1} I_\nu^{(j-1)}(x_k)]}{\Delta(\{x_k^2\})}.
\ee
This results was first obtained in \cite{brower-rossi-tan} 
and independently for equal masses in \cite{bars-green,kogut-snow}.
Using identities such as
\be
x^2 \del_x^2 &=& (x\del_x)^2 - x\del_x,\quad
x^3\del_x^3 = (x\del_x)^3 - 3 (\del_x)^2 + 2x\del_x, 
\label{derid}
\ee
we can rewrite this partition function in terms of
derivatives $\delta_k \equiv x_k \del_{x_k}$ as
\be
Z_{\nu,N_f}(M) = \frac {\det [\delta_k^{j-1} I_\nu(x_k)]}
{\Delta (\{x_k^2\})}.
\label{zdifm}
\ee
This form  of the partition function is also known as a 
$\tau$-function \cite{mmm,ake-dam-2000}.

\subsubsection{QCD Partition Function for Equal Masses}

The limit of equal masses in the partition function (\ref{zdifm}) 
can be obtained by writing
\be
x_k = x(1+\delta x_k),
\ee
and taking the limit $\delta x_k \to 0$.
Because all columns are the same for $\delta x_k = 0$, we have to 
expand the matrix elements to order $(\delta x_k)^{N_f -1}$. The expansion
to this order can be combined into
\be
\det[\delta^{j-1}_k I_\nu(x_k)]  
= \det[ x^{l-1} (\delta_x^{j-1}I_\nu(x))^{(l-1)}]
\det [\frac{(\delta x_k)^{l-1}}{(l-1)!}].
\ee
Using that
\be
\det [ (\delta x_k)^{l-1}] &=& \Delta(\{\delta x_k\}),\quad
\Delta(\{x_k^2\}) = x^{N_f(N_f-1)} \Delta(\{\delta x_k\}),\hspace*{0.3cm}
\ee
we obtain the partition function 
\be
Z_{\nu, N_f}(x) =c x ^{-N_f(N_f-1)}\det [ x^{l-1} 
((x\del_x)^{j-1}I_\nu(x) )^{(l-1)} ].
\ee
With the help of the identities (\ref{derid}) the derivatives can be
combined into derivatives $(x\del_x)^p$ resulting in
\be
Z_{\nu,N_f}(x) =c x ^{-N_f(N_f-1)}\det [ 
(x\del_x)^{l+j-2}I_\nu(x)  ]_{1\le j,l\le N_f}.
\label{zeqm}
\ee

\subsection{Toda Lattice Equation}

In this section we show that the QCD partition function for equal
masses satisfies a Toda lattice equation. This result and 
generalizations thereof were first obtained in \cite{Kharchev}.
The Toda lattice was originally introduced as
a one dimensional lattice
in which neighboring atoms interact via a potential that depends
exponentially on the distance. The Hamiltonian equations of motion
of this system can be written in the form of the Toda lattice equation
discussed below. Because of the existence of a Lax pair, they have
infinitely many constants of motion.
For a more elaborate discussion of the Toda lattice equation and
the relation to integrable systems, we refer to 
\cite{ForresterBook,dunne,tracy}.
Several subsections below are based on 
the paper by Forrester and Witte \cite{FW}.

\subsubsection{The Sylvester Identity}

We all know how to expand a the determinant 
matrix with respect to its co-factors given by
\be
C_{ij} = \frac \del {\del{ A_{ij}}} \det A.
\ee
What is less known is that there exists a remarkable identity
that relates co-factors to the double co-factors defined by
\be
C_{ij;pq} = \frac {\del^2}{\del{ A_{ij}} \del{A_{pq}}} \det A.
\ee
This identity, which is known as the Sylvester identity \cite{Sylvester}, 
is given
by
\be
C_{ij}C_{pq} -C_{iq} C_{pj} = \det A C_{ij;pq}.
\label{sylvester}
\ee
For example, it holds for a $2\times 2$ matrix with $i=j=1$ and
$p=q=2$.

\subsubsection{Toda Lattice Equation}

We apply the Sylvester identity to the determinant that appears
in the partition function (\ref{zeqm}).
For $i=j=N_f -1 $ and $p=q =N_f$ we obtain
\be
C_{N_f -1,N_f -1}C_{N_f, N_f} -C_{N_f -1, N_f} C_{N_f, N_f -1} = 
\det A C_{N_f-1, N_f-1;N_f,N_f},\hspace*{0.3cm}
\label{sylnf}
\ee
with matrix 
$A$ given by
\be
 A_{jl} \equiv (x\del_x)^{l+j-2}I_\nu(x).
\label{amat}
\ee
The derivative of a determinant is equal to the sum of determinants
with one of the rows replaced by its derivatives, or is equal to
sum of the determinants with one of the columns replaced by
its derivative (in both cases we have in total $N_f$ terms).
For the  matrix $A$ in (\ref{amat}) only differentiating the last row or 
column gives 
a nonzero result. This allows us to rewrite the co-factors as
derivatives of $\det A$. In particular, we find
\be
C_{N_f -1, N_f} &=& -x\del_x \det A_{N_f-1},\qquad
C_{N_f, N_f-1} = -x\del_x \det A_{N_f-1},\nn\\
C_{N_f , N_f} &=& \det A_{N_f-1},\qquad
C_{N_f -1, N_f-1} = (x\del_x)^2 \det A_{N_f-1}.\nn\\
C_{N_f -1, N_f -1; N_f, N_f} &=& \det A_{N_f-2}. 
\ee
To obtain the second
last identity we first differentiate the columns and then the rows.
Inserting this in (\ref{sylnf}) we find
\be
 (x\del_x)^2 \log \det A_{N_f-1} = \frac{\det A_{N_f} \det A_{N_f-2}}
{{\det}^2 A_{N_f-1}}.
\label{pretoda}
\ee
Next we substitute the relation between $\det A_{N_f}$ and the 
partition function
\be
\det A_{N_f} = \frac 1c x^{N_f(N_f-1)} Z_\nu^{N_f}(x).
\ee
The prefactor contributes a factor $x^2$ to the r.h.s. of 
(\ref{pretoda}) but does not contribute to its l.h.s.. After raising
$N_f$ by 1 we obtain the celebrated Toda lattice equation
\cite{mmm,ake-dam-2000}
\be
(x\del_x)^2 \log Z_{\nu,N_f}(x) = c x^2 \frac{Z_{\nu,N_f+1}(x)
Z_{\nu,N_f-1} (x)}{[Z_{\nu,N_f}(x)]^2} .
\label{toda}
\ee
The large-$x$ limit of the partition function for equal masses,
\be
Z_{\nu,N_f} (x) = \int_{U \in U(N_f)} dU {\det}^\nu U 
e^{\frac x2 {\rm Tr} (U+ U^{-1})},
\label{zlow}
\ee
is obtained by performing the $U$-integral
by a saddle point approximation including
the Gaussian fluctuations. The asymptotic behavior is given by
\be
Z_{\nu,N_f}(x) \sim \frac {e^{N_f x}}{x^{N_f^2/2}}.
\label{zasym}
\ee
Using this result to normalize the partition 
function we find that
$
c = N_f.
$

\subsection{Painlev$\acute {\rm e}$ System}

In this subsection we derive the Toda lattice
from the Backlund transformation of a Painlev$\acute {\rm e}$ 
equation \cite{FW}.

The partition function (\ref{zlow})
can be obtained from the ``double'' scaling limit of the random matrix
partition function
\be
Z_{\nu,N_f}^{\rm RMT}(x) = \int \prod_{k =1}^N [d\lambda_k \lambda_k^{2\nu 
+1}
(\lambda_k^2+m^2)^{N_f}] m^{\nu N_f}|\Delta(\{\lambda_l^2\})|^2
e^{-\frac {N\Sigma^2}2 \sum_l\lambda_l^2} ,\nn\\
\label{zranp}
\ee
where $x=mN \Sigma$ is kept fixed for
$N\to \infty$.
Using $x_k^2 = \lambda_k^2 \Sigma^2 N^2 +x^2$ as new
integration variables, we obtain
in this limit 
\be
Z_{\nu,N_f}^{\rm RMT}(x) 
&=& \prod_k[ \int_{x^2}^\infty  d x_k^2 (x_k^2-x^2)^\nu
x_k^{2N_f}] x^{\nu N_f} e^{x^2/4} e^{-\sum_k x_k^2/2N}
\nn\\ &\equiv&   x^{\nu N_f} e^{x^2/4}E_N([0,s=x^2], N_f,\nu).
\label{zgap}
\ee
$E_N([0,s=x^2], N_f,\nu)$ can be interpreted as the probability that there 
are no eigenvalues in the interval $[0,x^2]$ for
the joint probability distribution given
by the integrand of (\ref{zgap}). 
For $\nu=0$ this is the probability 
of the partition function with topological charge $N_f$ 
and no flavors.
If we introduce $\sigma(t)$ by
\be
E_N([0,s], N_f,\nu) = e^{-\int_0^s \frac {dt}t (\sigma(t) 
+ \frac 12\nu(N_f+\nu))},
\ee
then the function $\sigma(t)$ satisfies the Painlev$\acute {\rm e}$ 
equation \cite{FW},
\be
(t\sigma'')^2 -(N_f^2-\nu^2)(\sigma')^2 + \sigma'(4\sigma'-1)(\sigma-t\sigma') -
\frac{\nu^2}{16} =0.
\label{p3}
\ee
The boundary conditions for this differential equation follow from the
asymptotic behavior of the partition function (\ref{zlow}).
Using (\ref{zasym}) we find for the large $x$ behavior of 
$E_N([0,s=x^2,N_f,\nu)$,
\be
E_N([0,s=x^2,N_f,\nu) \sim x^{-N_f^2/2 -\nu N_f} e^{N_f x - x^2/4},
\ee
so that the large-$s$ behavior of $\sigma(s)$ is given by
\be
\sigma(s) \sim \frac{N_f}2 \sqrt s +\frac s4 -\frac{\nu^2}2 + 
\frac{N_f^2}4.
\label{sigasym}
\ee

The Painlev$\acute {\rm e}$ equation  (\ref{p3}) can be derived from 
the Hamiltonian \cite{FW}
\be
tH = q^2p^2 -(q^2+(N_f+\nu) q -t)p + N_f q
\ee
with the identification
\be
\sigma(t) = -\left .(tH)\right|_{t\to \frac t4} -\frac 12 (N_f+\nu )\nu
+\frac t4.
\ee
For example, we have the Hamiltonian equations of motion
\be
(tH)' &=& p,\quad
(tH)''=  p' = -\frac{\del H}{\del q},\quad
tq' = \frac {\del tH}{\del p}.
\label{hamilton}
\ee
Such Hamiltonians play an important role in the theory of exactly solvable
models. Hamiltonians with different values of $\nu$ and $N_f$ are connected
by a Backlund transformation. This is a canonical transformation
together with $(\nu, N_f)\to (\bar \nu , \bar N_f)$ such that, in 
the new variables, the same Painlev$\acute {\rm e}$ 
equation is satisfied. In our case,
we have the Backlund transformation (at fixed $\nu$)
\be
T: \hspace{1cm} N_f &\to& N_f +1,\nn \\
                H_{N_f} &\to& H_{N_f +1} = H_{N_f}
                       + q_{N_f} - q_{N_f}p_{N_f},\nn\\
               q_{N_f} &\to & q_{N_f+1},\nn\\
               p_{N_f} &\to & p_{N_f+1}.
\ee
Below we do not need the explicit transformation rules of $q_{N_f}$ and
$p_{N_f}$, but of the inverse transformation of $ q_{N_f}(p_{N_f}-1)$ which
is given by
\be
T^{-1} q_{N_f}(p_{N_f}-1) \to -q_{N_f}(p_{N_f}-1)+N_f +\nu 
-\frac{N_f}{p_{N_f}}.
\label{backtm}
\ee

If we define the $\tau$-function by
\be
\tau_{N_f} = e^{\int_0^t H_{N_f} dt},
\ee
one can easily derive the equalities
\be
t\del_t \log \frac{\tau_{N_f -1} \tau_{N_f+1}}{\tau_{N_f}^2}&=&
 tH_{N_f-1} -tH_{N_f}+tH_{N_f+1} -tH_{N_f},\nn\\
&=&-T^{-1}(q_{N_f}(1-p_{N_f})+(q_{N_f}(1-p_{N_f}),\nn\\
&=&t\del_t \log \del_t[tH_{N_f}],\nn\\
&=& t\del_t \log \del_t[t\del_t \log \tau_{N_f}].
\ee
To derive the second last equality we have used the inverse Backlund
transformation and the Hamilton equations (\ref{hamilton}).
Integrating this equation once and putting  the integration 
constant equal to zero we find the Toda lattice equation
\be
(t\del_t)^2 \log \tau_{N_f} = t \frac{\tau_{N_f+1} \tau_{N_f-1}}
{\tau^2_{N_f}}.
\ee 

\subsubsection{Solutions of the Painlev$\acute {\rm e}$ equation}

The probability $ E([0,s],N_f,\nu)$ is related to the partition
function (\ref{zlow}) by
\be
E([0,s],N_f,\nu) = s^{-N_f\nu/2} e^{-s/4} Z_{\nu,N_f}(\sqrt s).
\ee
For $N_f =0$ the partition function is normalized to 1 so that
\be
\sigma_{N_f =0}(s)= \frac s4 -\frac{\nu^2}2.
\ee  
Indeed this is a solution of the $PIII$ Painlev$\acute {\rm e}$ 
equation (\ref{p3}).

For $N_f = 1$ we have that
\be
E([0,s],N_f=1,\nu) = s^{-\nu/2} e^{-s/4} I_{\nu}(\sqrt s)
=
e^{-\int_0^s \frac {dt}t (\sigma(t) + \frac 12\nu(1+\nu))},\hspace*{0.5cm}
\label{painnf1}
\ee
resulting in another solution of (\ref{p3}):
\be
\sigma_{N_f =1}(s) = \frac s4 -\frac{\nu^2}2 -s\frac d{ds} \log I_\nu(\sqrt s).
\ee
For $x \to 0$ the modified Bessel function behaves as $I_\nu(x) \sim x^\nu$
so that the $t$-integral in (\ref{painnf1}) is well-behaved for $t\to 0$.
Only recursion relations for Bessel functions are required to 
show that (\ref{painnf1}) is a solution of the Painlev$\acute {\rm e}$ 
equation. 
Since $(-1)^\nu K_\nu(x)$ satisfies the same recursion relations as $I_\nu(x)$ 
this provides us with another solution of  the Painlev$\acute {\rm e}$ 
equation.
This solution corresponds to the partition function with $N_f = -1$
where
\be
E([0,s],N_f=-1,\nu) = s^{\nu/2} e^{-s/4} K_{\nu}(\sqrt s)
=
e^{-\int_0^s \frac {dt}t (\sigma(t) + \frac 12\nu(-1+\nu))},
\hspace*{0.5cm}
\label{pain1}
\ee
and satisfies the boundary condition for $N_f = -1$.

\subsubsection{The Bosonic Partition Function}
\label{intqc2}
 
The natural interpretation of $N_f = -1$ is as a
bosonic flavor. In this section we will derive the low energy limit
of the QCD partition function for $N_f$ bosonic flavors with equal masses. 
We already have seen in section \ref{symme3} that the Goldstone manifold for
$n$ bosonic quarks is given by $Gl(n) /U(n)$. Using the same invariance
arguments as before one obtains the low-energy effective partition function
\be 
Z_{\nu,-n} = \int_{Q \in Gl(n)/U(n)} {\det}^\nu (Q) 
e^{\frac 12 V\Sigma {\rm Tr}\,
M (Q + Q^{-1})}.
\label{zeffnu2}
\ee
In this case $Q$ can be diagonalized as 
$
Q = U {\rm diag}(e^{s_k}) U^{-1},
$
so that an eigenvalue representation of this partition function is
given by \cite{DV1}
\be
\int \prod_k ds_k \prod_k e^{\nu s_k} 
\prod_{k<l} (e^{s_k} - e^{s_l})(e^{-s_k} - e^{-s_l})
e^{x\sum_k \cosh s_k}.
\ee
The Vandermonde determinant can be written as
\be
\prod_{k<l} (e^{s_k} - e^{s_l}) = \det 
[ e^{p s_q}]_{0\le p\le n-1, \,1\le q\le n}
\ee
and a similar expression for $s_k \to -s_k$.
By expanding the two determinants the integrals can be written as
modified
Bessel functions which can be combined into a determinant as follows
\cite{DV1}
\be
Z_{\nu,-n}(x) =c_{-n}\det[K_{\nu+k+l}(x)]_{0\le k,l \le n-1}.
\label{zbos1}
\ee
From the observation that $(-1)^\nu K_\nu(x)$ and
$I_\nu(x)$ satisfy the same recursion relations, and that the factor
$(-1)^\nu$ does not affect the determinant, (\ref{zbos1})
can be rewritten as the $\tau$-function
\be 
Z_{\nu,-n}(x) = 
\frac{c_{-n}}{x^{n(n-1)}}\det[(x\del_x)^{k+1}Z_{\nu,-1}(x)]_{0\le 
k,l \le n-1}.
\label{bostau}
\ee
with
$
Z_{\nu,-1}(x) = K_\nu(x).
$
The bosonic partition function can also be analyzed along the same
lines as the fermionic  partition function. On the other hand this derivation
can be simply modified to obtain the partition function for $N_f$ fermionic 
flavors with equal mass. 

The bosonic partition functions thus satisfy the same Toda lattice equations
as the fermionic partition functions.
The semi-infinite hierarchies are connected by
\be 
\lim_{n \to 0}\frac 1n (x\del_x)^2 \log Z_{\nu,n}(x),
\ee
which  is related to a derivative of the resolvent.

\subsection{Replica Limit of the Toda Lattice Equation}
\label{intqc3}

The resolvent can be obtained from the replica limit 
\be
G(z) = \lim_{n \to 0} \frac 1{n} \log Z_{\nu,n}(z).
\ee
If we take the replica limit of the fermionic ($n>0$) or 
bosonic $(n<0)$
partition functions directly, we will obtain a result that differs from the
supersymmetric calculation. These problems can be avoided if the take 
the replica limit of the Toda lattice equation (\ref{toda}). 
With the normalization
$Z_{\nu,0}(x) = 1$ we obtain the relation
\be
x\del_x x G(x) = 2 x^2 Z_{\nu,1}(x) Z_{\nu,-1} (x).
\ee
Inserting the expressions for $Z_{\nu,1}$ and $Z_{\nu,-1}$ we find
\cite{SplitVerb1}
\be
G(x) =\frac \nu x + x(K_\nu(x)I_\nu(x) + K_{\nu-1}(x) I_{\nu+1}(x)),
\ee
which agrees with the result obtained from the 
supersymmetric method (\ref{val}).
This result
has also been derived from the solution of the Painlev$\acute {\rm e}$ equation
(\ref{p3}) for $n \to 0$ \cite{kanzieper}. 

The validity of the replica limit of the Toda lattice equation can
be proved by
extending the Toda lattice hierarchy to include an additional
spectator boson with mass $y$ and using the identity \cite{SplitVerb3}
\be
\lim_{n  \to 0}\frac 1n (x\del_x)^2 \log Z_{\nu,n}(x)= 
\lim_{y\to x} x\del_x(x\del_x + y \del_y) \log Z_{\nu,1,-1}(x,y).
\hspace*{0.5cm}
\ee

\subsection{Replica Limit for the GUE Two-Point Function}
\label{twopo1}

We have two possibilities for the generating function of the two-point
function of the Gaussian Unitary Ensemble: a fermionic
generating function or a bosonic generating function.
The fermionic (bosonic)
generating function for the two-point function is defined
by
\be
Z_n(x,y) = \int dH P(H) {\det}^n (x+i\epsilon +H)
{\det}^n (y-i\epsilon +H),
\label{zfer}
\ee
with $n > 0$ $(n<0)$.
We will consider the microscopic 
limit where 
$\pi(x-y)N \rho(x)\equiv r$ is kept fixed for $ N \to \infty$. In that
case the two-point function in the center of the spectrum only
depends on $r$ and is given by
\be
R_2(r)= - \lim_{n \to 0} \frac 1{n^2} \del_r^2Z_n(r),
\ee
both in the fermionic and the bosonic  case. 
 In an eigenvalue representation of the
Goldstone fields the 
microscopic limit of the generating function $Z_n(r)$ can be written as
\cite{critique,mex}
\be
Z_n(r) = \int_{-1}^1 \prod_kd u_k \prod_{k<l}(u_k-u_l)^2  
e^{ir\sum_k u_k}.
\label{partferm2}
\ee
This partition (\ref{partferm2}) can be written as a $\tau$-function. The first
step is to expand the Vandermonde determinant
\be
Z_n(r) =\int_{-1}^1 \prod_kd u_k \sum_{\sigma\pi}sg(\sigma\pi)
u_1^{\sigma(1) +\pi(1)} \cdots u_n^{\sigma(n) +\pi(n)}   
e^{ir \sum_k u_k}.
\ee
Next we use that
\be
\int_{-1}^1 d u_k u_k^a e^{ir u_k} = (\del_{ir})^a Z_1(r),
\ee
which results in \cite{kanzieper2001} 
\be 
Z_n(r) = n! [\det (\del_{ir})^{i+j} Z_1(x)]_{0 \le i,j \le n-1}.
\ee
The partition function $Z_1(x)$ is given by
\be
Z_1(r)= \int_{-1}^1 du e^{iru}.
\ee

The microscopic limit of 
bosonic partition function can be rewritten similarly.
The main difference is 
the convergence requirements of the bosonic 
integrals which are essential for the structure of the Goldstone manifold.  
In an eigenvalue representation of the Goldstone fields we find
\cite{critique}
\be
Z_{-n}(r) = \int_{1}^\infty \prod_k d u_k \prod_{k<l}(u_k-u_l)^2  
e^{ir \sum_k u_k}.
\label{zbos}
\ee
This partition function can also be written as a $\tau$-function.
By expanding the Vandermonde determinant we can express
this generating function as a determinant of derivatives
\be
Z_{-n}(r) = n! [\det (\del_{ir})^{i+j} Z_{-1}]_{0 \le i,j \le n-1},
\ee
with $Z_{-1}(r)$ given by
\be
Z_{-1}(r)= \int_{1}^\infty due^{iru}.
\ee

Because of the derivative structure of
the partition function, we can again use the Sylvester identity to
derive a Toda lattice equation. In this case we find
\be
\del_{ir}^2 \log Z_n(x) = n^2 \frac {Z_{n+1}(r) Z_{n-1}(r)}{[Z_n(r)]^2},
\label{todague}
\ee
where the factor $n^2$ follows from the choice of the
normalization constants. We have made this choice because the left hand
side is proportional to $n^2$. 
The two-point correlation is given by the replica limit of (\ref{todague})
\be 
R_2(r) &=& -\lim_{n\to 0}\frac 1{n^2} \del_r^2 \log Z_n(r)
= Z_1 (r) Z_{-1} (r)\nn \\  
&=&
\int_{-1}^1 du e^{iux} \int_1^\infty e^{iux}
 =  2i \frac{\sin x}x \frac{ e^{ix}}x,
\ee
which is the correct analytical result for 
the two-point function. This derivation explains the factorization
of the two-point function into a compact and a non-compact integral 
which characterizes the result obtained
by a supersymmetric calculation  \cite{Efetov}.
The fermionic partition functions, the bosonic partition functions and 
the super-symmetric partition function form a single integrable
hierarchy which are related by the Toda lattice equation \cite{SplitVerb1}.
A closely related way to derive the two-point function of the GUE is
to take the replica limit of the corresponding Painlev$\acute {\rm e}$ 
equation.
For a discussion of this approach we refer to \cite{kanzieper} which preceded
our work \cite{SplitVerb1} on the Toda lattice.

\section{QCD at Finite Baryon Density}
\label{qcmu}
In this Chapter we study the quenched microscopic spectrum of the QCD Dirac
operator at nonzero chemical potential when the Dirac operator
is non-Hermitian with eigenvalues scattered in the complex plane.
Using the replica limit of the Toda
lattice equation we obtain the exact analytical 
result for the microscopic spectral density \cite{SplitVerb2}.

\subsection{General Remarks}
\label{qcmu1}

The average spectral density of a non-Hermitian operator is given by
\be 
\rho(\lambda) = \langle \sum_k \delta^2(\lambda- \lambda_k) \rangle,
\ee
and the average resolvent is defined by
\be
G(z) = \left \langle\sum_k \frac 1{i\lambda_k +z}\right \rangle.
\ee
Using that $\del_{z^*} (1/z) = \pi \delta^2 (z)$ we easily derive
\be
\del_{z^*} G(z)|_{z=\lambda} = \pi \rho(\lambda).
\ee
The resolvent can be interpreted as the electric field in the plane
at point $z$ from charges located at the position of the eigenvalues.
For example, Gauss law is given by
\be
\oint_C G(z) d z = 2\pi i Q,
\ee
where $Q$ is the number of eigenvalues enclosed by $C$.

\subsection{The Ginibre Ensemble}

The obtain a better understanding of the resolvent for
a non-Hermitian random matrix ensemble, we first consider
the Ginibre ensemble
\cite{ginibre} defined by the probability distribution
\be
\rho(C) = e^{-N{\rm Tr} C C^\dagger},
\ee
with $C$ a complex $N\times N$ matrix. 
The eigenvalues of $C$ are given
by the solutions of the secular equation
$
\det (C- \lambda_k) = 0.
$
If all eigenvalues are different, the matrix $C$ can  be decomposed as
\be
C = V \Lambda V^{-1},
\ee
where $V$ is a similarity transformation and 
$\Lambda ={\rm diag}(\lambda_1,\cdots, \lambda_N)$. The joint eigenvalue
distribution is obtained by using the decomposition
\be
C = U T U^{-1},
\ee
with $U$ a unitary matrix and $T$ an triangular matrix with the
eigenvalues of $C$ on the diagonal. After integrating out the upper 
triangular matrix elements we obtain
(see the lectures by Zabrodin \cite{zabrodin} for a derivation), 
\be 
\rho(\{\lambda_k\}) = | \Delta(\{ \lambda_k\}) |^2 e^{-N\sum_k |\lambda_k|^2}.
\ee
This distribution can be interpreted as repulsive charges in the plane balanced
by an external force $N|z|$. 
The resolvent is equal to the electric field of the eigenvalues. 
For an equilibrium distribution we have $G(z) = N|z|$.
Using that the eigenvalue density is 
spherically symmetric, we find from Gauss law
\be
2\pi r |G| = 2\pi \int_0^r \rho(r') dr',
\ee
so that 
$
\rho(r) = N/\pi.
$
Because the total number of eigenvalues is equal to $N$, 
they are located inside the circle
$|z| =1$. The resolvent is thus given by
\be
G(z) &=& Nz^*\theta(1-|z|) + \frac Nz \theta(|z|-1).
\ee

\subsection{QCD at Nonzero Chemical Potential}
\label{munonzero}

The QCD partition function at nonzero chemical potential $\mu$ is given by
\be
Z_{\rm QCD} =\sum_k e^{-\beta (E_k - \mu N_k)},
\ee
where $E_k$ is the energy of the state, and $N_k$ is the quark number of
the state. At zero temperature ($\beta \to \infty$) the partition function
does not depend on $\mu$ for $\mu < m_N/N_N$, where $N$ is the particle
with the smallest value of $m_N/N_N$. For QCD $N$ is the nucleon with
quark number $N_N = 3$. This implies that the chiral 
condensate does not depend on  $\mu$ for $\mu < m_N/N_N$.

The QCD partition function can be written as a Euclidean path integral
with the fermionic part of the Lagrangian density given by
\be
{\cal L} = \bar \psi D \psi + m\bar \psi \psi + \mu \bar \psi \gamma_0 \psi.
\ee
with $D$ the anti-Hermitian Dirac operator. Since $\mu \gamma_0$ is
Hermitian, the Dirac operator as a whole is non-Hermitian. As a consequence,
the eigenvalues are scattered in the 
complex plane \cite{all}. The fermion determinant
is in general complex. This means that it is not possible to study
the QCD partition function by means stochastic methods  which severely limits
our knowledge of QCD at nonzero chemical potential.

The question we wish to address is if there is a domain where
the fluctuations of the Dirac eigenvalues are universal and
can be obtained from a random matrix
partition function with the global symmetries QCD, or equivalently from a
 chiral Lagrangian.  In this domain we will calculate the resolvent
and the spectral density
from the replica limit of the Toda lattice equation
\cite{SplitVerb2,AOSV}. 

\subsubsection{Generating Function for the Quenched Spectral Density}

The quenched spectral density is given
by the replica limit \cite{Girko,Goksch,misha}
\be
\rho^{\rm quen}(z,z^*,\mu) 
= \lim_{n \to 0} \frac 1{\pi n}\del_z \del_{z^*} \log 
Z_{\nu,n}(z,z^*,\mu),
\ee
with generating function given by 
\be
Z_{\nu,n}(z,z^*,\mu) = \langle {\det}^n(D + \mu\gamma_0 +z)
{\det}^n(-D + \mu\gamma_0 +z^*)\rangle. 
\label{zngen}
\ee
The product of the determinants in (\ref{zngen}) 
can be written as the determinant of \cite{Feinnh,TV}
\be
 \left( \begin{array}{cccc} id+\mu &0 &z& 0 \\
                            0& id-\mu & 0& z^*\\
                 z & 0 & id^\dagger+\mu  & 0 \\
                  0 & z^* & 0& id^\dagger -\mu \\ \end{array} \right )
\equiv \mat id +\mu_1& M_{RL}\\M_{LR} & id^\dagger + \mu_2 \emat.
\nonumber \\
\ee
Here we have  used the decomposition of the Dirac operator given
in (\ref{diracstructure}).
We observe that the $U(2n) \times U(2n)$ flavor symmetry is broken
by the chemical potential term and the mass term. Invariance is recovered
by transforming the mass term as in the case of zero chemical potential
(see (\ref{umass}))
and the chemical potential term by a local gauge transformation
\cite{KST}.  For the chemical potential matrices the latter
transformation is simply given by
\be
\mu_1 \to U_R \mu_1 U_R^{-1}, \qquad \mu_2 \to U_L \mu_2 U_L^{-1}.
\ee
The low-energy limit of quenched QCD should have the same transformation
properties.  In the domain
$\mu \ll 1/{L}$ and $z\Sigma \ll F^2/L^2$, 
we only have to consider the zero momentum modes. 
Using that the Goldstone fields transform as $U \to U_R U U_L^{-1}$ 
we can write down the following invariants to first order in the
quark mass and to second order in the chemical potential
\be
{\rm Tr}\mu_k^2, \quad {\rm Tr} U^{-1}\mu_1 U \mu_2, \quad
{\rm Tr} M_{RL} U^{-1}, \qquad {\rm Tr} M_{LR} U.
\ee
The low energy effective partition function is therefore given by
\be
Z_{\nu,n}(z,z^*,\mu) = 
\int_{U(2n)} dU {\det}^\nu U \,e^{-\frac{F^2\mu^2 V}4 {\rm Tr}[U,B][U^{-1},B]
+\frac {\Sigma V}2{\rm Tr}M(U+U^{-1})},\hspace*{0.3cm}
\label{genmu}
\ee
where
\be
B= \mat {\bf 1}_n &0 \\ 0 & - {\bf 1}_n \emat, \qquad
M = \mat z{\bf 1}_n &0 \\ 0 & z^* {\bf 1}_n \emat.
\ee

\subsubsection{Random Matrix Model}

The partition function (\ref{genmu}) can be obtained from the large $N$ limit
of a random matrix
model with the global symmetries of the QCD partition function.  
For $\nu=0$, the model is defined by an integral over
$N/2\times N/2$ complex matrices
\cite{SV,JV,misha},
\be
Z_{\nu=0}(m_f, \mu) = \int dW  \prod_{f=1}^{N_f} \det (D(\mu) + m_f) 
e^{-\frac N2 \Sigma^2{\rm Tr} W W^\dagger}.
\ee
The Dirac matrix has the structure
\be
D(\mu) = \mat 0 & iW +\mu \\ iW^\dagger +\mu & 0 \emat.
\label{rmtmisha}
\ee
For QCD with three or more colors in the fundamental representation, the
matrix $W$ is complex ($\beta_D =2$). One can
also introduce random matrix ensembles with $\beta_D = 1$ or $\beta_D =4$
by choosing the matrix elements of $W$ real or quaternion real,
respectively \cite{HOV}.

An alternative random matrix model \cite{james} is obtained by replacing the 
identity  matrix  that multiplies $\mu$ by  a complex matrix
with the same distribution as $W$. This random matrix model
is in the same universality class but turns out to be mathematically
simpler. In particular, the joint eigenvalue
distribution has been derived \cite{james} which makes it possible to 
calculate correlation functions by means of the
orthogonal polynomial method.

\subsubsection{Mean Field Analysis}
\label{mftmu}

The macroscopic spectral density of the partition function (\ref{genmu})
can be easily obtained by means of  a saddle point 
approximation \cite{TV}.
Using an Ansatz that is diagonal in replica space, 
\be
U = \mat \cos \theta & e^{i\phi} \sin \theta \\ -e^{-i\phi} \sin \theta &
\cos \theta \emat,\hspace*{0.3cm}
\ee
the partition function is given by
\be
Z_{n}(z,z^*,\mu) = e^{nV[2\mu^2 F^2 \sin^2\theta + \Sigma(z+z^*)\cos\theta 
]}.
\ee
The extrema are at
\be
\cos \theta = 1, \quad {\rm or} \quad 
\cos \theta =\frac{\Sigma(z+z^*)}{4F^2 \mu^2}.
\ee
The critical value of $\mu$ is at the point where the two saddle points
coincide,
\be
\mu_c^2 = \frac {\Sigma|z+z^*|}{4F^2}.
\ee
The partition function at the saddle point is given by
\be
\mu  < \mu_c &:& Z_n(z,z^*,\mu) = e^{nV\Sigma(z+z^*)},\nn\\
\mu  > \mu_c &:& Z_n(z,z^*,\mu) = e^{nV(2\mu^2F^2 + 
\Sigma^2(z+z^*)^2 /(8F^2\mu^2) )}.
 \ee
For the resolvent and the spectral density we thus find
\be
\mu  < \mu_c &:& G^{\rm quen}(z,z^*,\mu) = V\Sigma, \quad \rho^{\rm 
quen}(z,z^*,\mu) = 0, \\
\mu  > \mu_c &:& G^{\rm quen}(z,z^*,\mu) = \frac 
{V\Sigma^2(z+z^*)}{4\mu^2F^2},
\quad \rho^{\rm quen}(z,z^*,\mu)= \frac{\Sigma^2 V}{4\mu^2 F^2}.
\hspace*{0.3cm}\nn \label{mft}
 \ee
The eigenvalues are located inside a strip of width $4F^2\mu^2/\Sigma$.

\subsection{The Microscopic Spectral Density}

The microscopic spectral density cannot be obtained from a mean 
field analysis. The assumption in the derivation of the
previous section is that saddle point is proportional to 
the identity in replica space,
so that the replica limit can be obtained
from the calculation with one replica. The
generating function for the microscopic spectral density depends
in a nontrivial way on the number of replicas which, as we have 
seen before, can be obtained from the replica limit of a
Toda lattice equation. In this subsection we closely follow
\cite{SplitVerb2}.
The starting point is a remarkable 
integration formula to be discussed next.

\subsubsection{Integration Formula}
\label{qcmu2}

By decomposing a $U(2n)$ matrix as
\be
U = \mat u_1 & \\ & u_2 \emat \mat v_1 & \\ & v_2 \emat  
\mat 
\sqrt{1-b^2}& b  \\ 
b & -\sqrt{1-b^2} \emat
\mat v_1^\dagger & \\ & v_2^\dagger \emat, 
\hspace*{0.3cm}\label{paramu}
\ee
with $u_1,\, u_2, \, v_1 \in U(n)$, $ v_2 \in U(n)/U^n(1)$ 
 and $b$ a diagonal matrix, 
the following integration formula can be proved \cite{SplitVerb2}
\be
&&\int_{U(2n)} dU {\det}^\nu U e ^{\frac 12 {\rm Tr} [ M(U + U^{-1}] + 
\sum_p a_p {\rm Tr}[(UBU^{-1} B)^p] } \nonumber\\ 
&=& \frac {c_n}{(xy)^{n(n-1)}} {\det} 
[ (x\del_x)^k (y\del_y)^l Z_{\nu,1}(x,y) ]_{
0\le k,l \le n-1},\hspace*{0.3cm}
\label{int}
\ee
where $c_n$ is an $n$-dependent constant and
\be
Z_{\nu,1}(x,y) = \int_0^1 \lambda d\lambda I_\nu(\lambda x) I_\nu(-\lambda 
y)
e^{2 \sum_p a_p \cos (2p\cos^{-1} \lambda)}.
\ee

\subsubsection{Toda Lattice Equation at Nonzero Chemical Potential}

Using the integration formula (\ref{int}) for $ p=1$ we find that the
zero momentum partition function $Z_{\nu,n}(z,z^*,\mu)$ (see Eq. 
(\ref{genmu}))
can be written as
\be
Z_{\nu,n}(z,z^*,\mu) = \frac{c_n}{(zz^*)^{n(n-1)}} \det[(z\del_z)^k 
(z^*\del_{z^*})^l 
Z_{\nu,1}(z,z^*)]_{0\le k,l\le n-1},
\label{mutau}
\ee
where
\be
Z_{\nu,1}(z,z^*,\mu) = \int_0^1 \lambda d\lambda e^{-2VF^2\mu^2 
(\lambda^2-1)}
|I_\nu(\lambda zV\Sigma)|^2.
\ee 
By applying the Sylvester identity to the determinant in (\ref{mutau})
for $i=j=n-1$ and $p=q=n$ and expressing the cofactors as derivatives,
we find a recursion relation that can
be written in the form of the Toda lattice equation
\be
z\del_z z^*\del_{z^*} \log Z_{\nu,n}(z,z^*,\mu) =
\frac{\pi n}2 (zz^*)^2 \frac{Z_{\nu,n+1}(z,z^*,\mu) 
Z_{\nu,n-1}(z,z^*,\mu)}
{[Z_{\nu,n}(z, z^*,\mu)]^2}.\hspace*{0.3cm} 
\ee
For the spectral density we find the simple expression
($Z_{\nu,0}(z, z^*,\mu)=1$)
\be
\rho^{\rm quen}(z,z^*,\mu) = \lim_{n \to 0} \frac 1{\pi n} \del_z 
\del_{z^*} \log
Z_{\nu,n}(z, z^*) =
\frac {zz^*}2 Z_{\nu,1}(z,z^*) Z_{\nu,-1}(z,z^*).\hspace*{0.3cm}\nn\\
\label{rhosimple}
\ee
What remains to be done is to calculate the bosonic partition function 
for $n=-1$ which will be completed in the
next subsections.

\subsubsection{The Bosonic Partition Function}
\label{qcmu4}

In this subsection we evaluate the low-energy limit of the QCD
partition function at nonzero chemical potential
for one bosonic quark and one conjugate bosonic quark and $\nu=0$. 
We closely follow \cite{SplitVerb2}. 
Because of convergence 
requirements, the inverse determinants of nonhermitian operators
have to be regulated. This is achieved by expressing them as the determinant
of a larger Hermitian operator \cite{Feinnh}
\be
&&{\det}^{-1} \mat z& id + \mu\\id^\dagger +\mu &z \emat   
{\det}^{-1} \mat z^*& -id + \mu\\-id^\dagger +\mu &z^* \emat  \nn\\
& = & \lim_{\epsilon \to 0}
{\det}^{-1} \matf \epsilon &       0  &               z &    id+\mu\\
      0                  & \epsilon & id^\dagger +\mu & z \\
      z^*                &-id + \mu & \epsilon        & 0 \\
      -id^\dagger+\mu        & z^*      & 0               & \epsilon \ematf
\\
&=&
\int
\exp[ i \sum_{j=1}^{N/2} \phi_k^{j\,*}
\matf \epsilon & z        &  id+\mu & 0\\
      z^*   &\epsilon   & 0& id - \mu         \\
    -id^\dagger +\mu&  0               & \epsilon &-z^*\\
      0 &  -id^\dagger -\mu &-z& \epsilon  \ematf_{kl}\phi_k^j ].\nn
\label{detzm2}
\ee 
The mass matrices are given by
\be
\zeta_1 = \mat \epsilon & z \\ z^* & \epsilon \emat 
\qquad{\rm and}\qquad
\zeta_2 = \mat \epsilon & -z^* \\ -z & \epsilon \emat =-I\zeta_1 I.
\label{masszeta}
\ee
with $I \equiv i\sigma_2$.
For the random matrix model (\ref{rmtmisha}) we have that $d=W$.
The Gaussian integral over $W$ results in the 4-boson  term
$
\exp  [ - 2 {\rm Tr} \overline{Q_1} \,\overline{Q_2}/N  ] 
$
with
\be
\overline{Q_1} \equiv \mat  \phi^{*}_1\cdot \phi_1
&\phi_1^* \cdot \phi_2\\
                  \phi_2^*\cdot \phi_1 & 
 \phi_2^* \cdot \phi_2 \emat ,
\qquad
\overline{Q_2} \equiv \mat \phi_3^*\cdot \phi_3 & \phi_3^*\cdot \phi_4\\
                   \phi_4^*\cdot \phi_3 & \phi_4^*\cdot \phi_4 \emat , 
\label{Q12}
\ee 
and we have used the notation
$
\phi^*_k \cdot \phi_l = \sum_{i=1}^{N/2} \phi_k^{i\,*} \phi_l^i.
$ 
Instead of the usual Hubbard-Stratonovitch transformation, we linearize
the 4-boson interaction term by the Hermitian matrix $\delta$ function
\be
\delta(Q_i -\overline{Q_i}) = \frac 1{(2\pi)^4}
\int dF e^{-i {\rm Tr}F(Q_i-\overline{Q_i})} , 
\ee 
where the integral is over Hermitian matrices $F$. We thus find the identity
\be
\exp \left [-\frac 2N {\rm Tr} \overline{Q_1}  \,\,\overline{Q_2}  \right ]
\sim
\int dQ_1dQ_2 \int dF dG e^{ {\rm Tr} [-i F(Q_1 -\overline{Q_1})
iG(Q_2 -\overline{Q_2}) -\frac 2N Q_1  Q_2  ]}. 
\nn \\
\ee
The integral over the $\phi_k$ is uniformly convergent in  
$F$ and $G$ which justifies
the interchange of the order of the integrals.
This results in the partition function
\be
Z_{-1} &=&
\int dQ_1dQ_2 \int dF dG e^{ {\rm Tr} [-i \frac{N}{2}FQ_1 -
i\frac{N}{2}GQ_2 +i \frac{N}{2}\zeta_1^T(Q_1 -IQ_2 I) -\frac{N}{2}  
Q_1  Q_2  ]}\nn \\
&&\times{\det}^{-\frac{N}{2}}
\mat \epsilon  + F & \mu\sigma_3 \\ 
\mu\sigma_3 & \epsilon + G  \emat,
\ee
where we have used a block notation and the mass matrices (\ref{masszeta}).
We have also simplified 
this integral by changing integration variables according
to
$F \to F-\zeta_1^T$ and $G\to G+I \zeta_1^TI$ and $Q_i \to N Q_i/2,
\, i= 1,2 $. 
For reasons of convergence we have kept the  infinitesimal increments
inside the determinant.
In the weak
nonhermiticity limit, where $\mu^2N$ is kept  fixed for $N \to \infty$, 
the determinant can be
approximated by  
\be 
{\det}^{-\frac{N}{2}}
\mat \epsilon  +F & \mu\sigma_3 \\ \mu\sigma_3 & \epsilon + G \emat
&=& {\det}^{-\frac{N}{2}} (\epsilon +F){\det}^{-\frac{N}{2}} (\epsilon + G)
\\ &&\hspace*{-1cm}\times
\exp\left [\frac{N\mu^2}{2} \frac 1{\epsilon +F}\sigma_3 
\frac 1 {\epsilon + G} \sigma_3\right ](1 +{\cal O}\left(\frac1N\right)).\nn
 \ee
The  $F$ and $G$ variables in the $\mu^2 N$ term
can be replaced the saddle point values of $F$ and $G$ at $\mu=0$
given by  $(\epsilon+F)Q_1 =i$ and $(\epsilon +G) Q_2 = i$.
The remaining integrals over $F$ and $G$ are Ingham-Siegel
integrals given by \cite{yan}
\be
\int dF {\det}^{-n} (\epsilon + F)  e^{i{\rm Tr}Q F}\sim
 \theta (Q){\det}^{n-p}(Q) e^{-i\epsilon\Tr Q},
\label{IS}
\ee
where the integral is over  $p \times p$ Hermitian matrices,
${\rm Im} \epsilon < 0$, and $\theta(Q)$
denotes that $Q$ is positive definite.
These manipulations result in
\be 
Z_{-1}(z,z^*,\mu) &=& 
\int dQ_1dQ_2 \theta(Q_1) \theta(Q_2) {\det}^{\frac{N}{2}-2}(Q_1Q_2)  
\\ &&\times
e^{ {\rm Tr }[
 i \frac{N}{2}\zeta_1^T(Q_1 -IQ_2I ) -\frac{N}{2} Q_1  Q_2   -\frac{N}{2}\mu^2
Q_1 \sigma_3 Q_2 \sigma_3 ]}. \nn
\ee 
In the limit $N\to \infty$ the integrals over the massive modes can
be performed by a saddle point approximation.
The 
saddle point equations are given by
\be
Q_1^{-1}-Q_2 = 0, \qquad  Q_2^{-1} -Q_1 = 0.
\ee
Both equations can be rewritten as
\be
Q_1 = Q_2^{-1},
\ee
and therefore only four of the modes, which we choose to be $Q_2$, 
can be integrated out by a 
saddle-point approximation. The quadratic fluctuations give rise to a factor
$\pi^2/{\det}^2 Q_1$.
The integral over the remaining modes has
to be performed exactly. We thus arrive at the partition function
\cite{SplitVerb2}
\be
Z_{-1}(z,z^*,\mu) = 
\int \frac{dQ_1}{{\det}^2 Q_1} \theta(Q_1) 
e^{ {\rm Tr} [ 
 i \frac{N}{2}\zeta_1^T(Q_1 -I Q_1^{-1}I )   -\frac{N}{2}\mu^2
Q_1 \sigma_3Q_1^{-1} \sigma_3 ]}. \hspace*{0.3cm}
\label{zminq}
\ee 
Before evaluating this integral, we rederive this
partition function based on the symmetries of the QCD partition function.

\subsubsection{Symmetries of $Z_{-1}(\mu)$}

For $\mu =0$ and $\zeta_1=\zeta_2=0$
the symmetry of the partition 
function (\ref{detzm2}) are the $Gl(2) \times Gl(2)$ transformations,
\be
\vect \phi_1 \\ \phi_2 \evect \to U_1  \vect \phi_1 \\ \phi_2 \evect, \qquad
\vect \phi_1^* \\ \phi_2^* \evect \to 
\vect \phi_1^* \\ \phi_2^* \evect U_2^{-1},\nn \\
\vect \phi_3 \\ \phi_4 \evect \to U_2  \vect \phi_3 \\ \phi_4 \evect, \qquad
\vect \phi_3^* \\ \phi_4^* \evect \to  
\vect \phi_3^* \\ \phi_4^* \evect U_1^{-1},
\ee
where we have disregarded convergence.
 This symmetry can be extended to 
nonzero mass or chemical potential if we adopt the transformation rules
\be
\zeta_1 \to U_2 \zeta_1 U_1^{-1}, &\qquad& \zeta_2 \to U_1 \zeta_2 U_2^{-1},
\nn \\ 
\mu_1 \to U_2 \mu_1 U_2^{-1}, &\qquad& \mu_2 \to U_1 \mu_2 U_1^{-1},
\ee
where $\mu_1$ is the chemical potential matrix that is added $id$ and
$\mu_2$ is the chemical potential matrix that is added to $-id^\dagger$.
These matrices are introduced for the sake of  discussing 
the transformation properties
of the partition function (\ref{detzm2}) and will ultimately be 
replaced by their original values
$
\mu_1 = \mu_2 = \mu \sigma_3.
$
The chiral symmetry is broken spontaneously to $Gl(2)$
by the chiral condensate. Because the 
bosonic integral has to converge, 
the Goldstone manifold is not $Gl(2)$ but rather
$Gl(2)/U(2)$, i.e. the coset of positive definite matrices
as in the case of zero chemical potential. Under 
a  $Gl(2)\times Gl(2)$ transformation the Goldstone fields transform
as
\be
Q \to U_1 Q U_2^{-1}.
\ee
The
low energy effective partition function should have the same 
transformation properties as the microscopic partition function (\ref{detzm2}).
To second order in $\mu$ and first order in the mass matrix we
can write down the following invariants
\be
{\rm Tr} \zeta_1 Q, \quad {\rm Tr} \zeta_2 Q^{-1}, \quad
{\rm Tr} Q\mu_1 Q^{-1} \mu_2 ,\quad {\rm Tr} \mu_1^2, \quad {\rm Tr}
\mu_2^2.   
\ee  
 We also have the discrete symmetry that the partition function is
invariant under the interchange of $\zeta_1$ and $\zeta_2$. This symmetry
implies that the coefficients of the two mass terms in the effective
partition function are the same. Using that the integration measure 
on positive definite Hermitian matrices is given by $dQ/{\det}^2 Q$,
we finally arrive at the effective
partition function
\be
Z_{\nu,-1}(z,z^*) = \int_{Q \in Gl(2)/U(2)} \frac{{\det}^\nu(Q)dQ}
{{\det}^2 (Q)}  
e^{-\frac{F^2\mu^2 V}4 {\rm Tr}[Q,B][Q^{-1},B]
+\frac {i\Sigma V}2{\rm Tr}(\zeta_1 Q+\zeta_2 Q^{-1})},\nonumber\\
\label{z-1} 
\ee
The partition function (\ref{zminq}) for $\nu=0$ is recovered after making
the identification $V\to N$, $\Sigma \to 1$ and $ F^2 \to 1$
and $Q\to Q^T$.

\subsubsection{Calculation of the Integral over $Q$}
\label{qint}
To evaluate the integral (\ref{z-1}) we use the parameterization
\be
Q = e^t \mat e^r \cosh s  & e^{i\theta}\sinh s \\
          e^{-i\theta}\sinh s &   e^{-r} \cosh s \emat.
\ee 
where
\be
r \in \langle -\infty, \infty \rangle, \quad
s \in \langle -\infty, \infty \rangle, \quad
t \in \langle -\infty, \infty \rangle, \quad
\theta \in \langle 0, \pi \rangle \quad.
\ee
The Jacobian relating the measures $dQ/{\rm det}^2Q$ and 
$dr ds dt d\theta$ is given by
\be
J = 4 e^{4t}\cosh s \sinh s.
\ee
We first perform the integral over $r$, which gives a factor 
$2K_0(2N\epsilon \cosh s \cosh t)$ with leading singularity given by 
$\sim -\log \epsilon$. This factor is absorbed in the normalization of 
the partition function. 
Then the integral over $\theta$ gives a
Bessel function. Introducing $u = \sinh s$ as new integration variable
we find \cite{SplitVerb2}
\be 
Z_{\nu,-1}(z,z^*,\mu) &=&
C_{-1}\int_{-\infty}^\infty dt  \int_0^\infty du e^{2\nu t}
J_0(2V u(x^2 \cosh^2 t
+y^2 \sinh^2 t)^{1/2})\nn\\&&\times
e^{-\mu^2 F^2 V (1+2 u^2)}.
\label{zminj}
\ee
To do the integral over $u$ we use the known integral
\be
\int_0^\infty dx x^{a+1} e^{-\alpha x^2} J_a(\beta x) = 
\frac{\beta^a}{(2\alpha)^{a+1}} e^{-\beta^2/4\alpha}.
\ee
This results in
\be
Z_{\nu,-1}(z,z^*,\mu) &=& \frac {C_{-1}e^{-V\mu^2 F^2} }{4\mu^2 F^2 V } 
\int_{-\infty}^\infty  dt   e^{2\nu t}
e^{-\frac{V (x^2 \cosh^2 t +y^2 \sinh^2 t)}{2\mu^2 F^2}}.\nn \\
\ee
Using that $\cosh^2 t = \frac 12 + \frac 12 \cosh 2t$ and
$\sinh^2 t = -\frac 12 + \frac 12 \cosh 2t$, the integral
over $t$ can be rewritten as a modified Bessel function
resulting in \cite{SplitVerb2}
\be
Z_{\nu,-1}(z=x+iy,z^*) = 
C_{-1} \,
e^{\frac{V\Sigma^2 (y^2-x^2)}{4 \mu^2F^2}}
K_\nu\left ( \frac {V\Sigma^2(x^2+y^2)}{4 \mu^2 F^2}\right ).
\label{z-1a}
\ee

\subsubsection{The Dirac Spectrum at Nonzero Chemical Potential}

The final result for the quenched spectral density is obtained 
by substituting the partition functions $Z_{\nu,1}^\nu(z,z^*,\mu)$
and $Z_{\nu,-1}(z=x+iy,z^*,\mu)$ in expression (\ref{rhosimple})
obtained from the replica limit of the Toda lattice equation.
We find,
\be
\rho^{\rm quen}(x,y,\mu) &=& \frac {V^3\Sigma^4}{2\pi F^2\mu^2}(x^2+y^2)
e^{\frac{V\Sigma^2 (y^2-x^2)}{4 \mu^2F^2}}
K_\nu\left ( \frac {V\Sigma^2(x^2+y^2)}{4 \mu^2 F^2}\right )
\nonumber \\ && \times
\int_0^1 \lambda d\lambda e^{-2VF^2\mu^2 \lambda^2)}
|I_\nu(\lambda zV\Sigma)|^2.
\label{final}
\ee 
The normalization constant has been chosen such that    the $\mu \to 0$  limit
of $\rho^{\rm quen}(x,y,\mu)$ for large $y$ is given by $\Sigma V/\pi$ 
(see below).

In the limit ${\rm Re}(z)\Sigma/\mu^2 F^2 \ll 1$ the upper limit of the 
integral in (\ref{final}) can be extended to infinity. Using the known
integral
\be
\int_0^\infty \lambda d\lambda e^{-2VF^2\mu^2 \lambda^2} |I_\nu(\lambda 
zV\Sigma)|^2 = \frac { e^{\frac{(z^2+z^{*\, 2}) \Sigma^2 V}
{8\mu^2 F^2}}}{4\mu^2 F^2 V} 
I_\nu\left(\frac{zz^* V\Sigma^2}{4\mu^2F^2}  \right ),\vspace*{0.3cm}
\ee
the spectral density can be expressed as
\be
\rho^{\rm quen}(x,y,\mu) = \frac {2}{\pi} u^2 zz^* K_\nu(zz^*u) 
I_\nu(zz^*u)
\quad {\rm with} \quad u = \frac{V \Sigma^2 }{4\mu^2 F^2}.
\ee
Therefore, the spectral density becomes a universal function that
only depends on a single parameter $u$. This parameter can be rewritten
in a more physical way as
$
u =  \pi  \rho^{\rm asym}(x,y,\mu).
$
For the dimensionless ratio we obtain 
\be
\frac{\rho^{\rm quen}(x,y,\mu)}{\rho_{\rm asym}(x,y,\mu)} = 
2 u zz^* K_\nu(zz^*u) I_\nu(zz^*u),
\ee
which is universal combination that depends only on a single universal
combination $zz^*u$. (This result was obtained in collaboration
with Tilo Wettig).

In the thermodynamic limit the Bessel functions can be approximated
by their asymptotic limit.
This results in
\be
\rho^{\rm quen}(x,y,\mu) &=& \frac {V^2\Sigma^2}{2\pi F\mu \sqrt{2\pi V}}
\int_0^1 d\lambda e^{-2VF^2\mu^2(\lambda - \frac{|x|\Sigma}{2F^2\mu^2})^2}.
\ee
For $V\to \infty$ the integral over $\lambda$ can be performed by a
saddle point approximation. If the saddle point is outside the
range $[0,1]$ the integral vanishes for $V\to \infty$. We thus find for
the spectral density
\be
\rho^{\rm quen}(x,y,\mu) = \frac {V\Sigma^2}{4\pi \mu^2 F^2 } \quad {\rm 
for} 
\quad |x |< \frac {2F^2 \mu^2}\Sigma
\ee
and $\rho^{\rm quen}(x,y,\mu) =0 $ outside this strip. This result is in 
agreement
with the mean field analysis \cite{TV} of the effective partition 
function given in section \ref{mftmu}.
For the integrated eigenvalue density we find
\be
\int_{-\infty}^\infty dx \rho^{\rm quen}(x,y,\mu) = \frac{\Sigma V}\pi
\ee
in agreement with the eigenvalue density at $\mu = 0$.

\section{Full QCD at Nonzero Chemical Potential}
\label{full}
In quenched QCD the chiral condensate  $G(m,\mu) \sim m/\mu^2$ in the
region where the eigenvalues are located. In section
\ref{munonzero} we have argued that
that chiral condensate in full QCD does not depend on $\mu$ for
$\mu < m_N /N_N$.
The conclusion is that the presence of the fermion determinant 
completely alters the vacuum structure of the theory. The question we
wish to address is how we can understand this based on the spectrum
of the QCD Dirac operator. For simplicity we only consider the case of
$N_f =1$ and $\nu=0$.  

The average spectral density in full QCD is defined by 
\be
\rho^{\rm full}(x,y,\mu) = \langle \sum_k \delta^2(x+iy-\lambda_k) 
\det (D+m +\mu \gamma_0) \rangle.
\ee
The low-energy limit of the generating function for the spectral density
can again be written as a $\tau$-function \cite{AOSV}. The spectral density is
then obtained from the replica limit of the corresponding Toda lattice
equation. 
The result is
\be
\rho^{\rm full}(x,y,\mu) &=& \frac {V^3(x^2+y^2)\Sigma^4}{2\pi\mu^2F^2}
e^{\frac{V(y^2-x^2)\Sigma^2}{4\mu^2F^2}}K_0(
\frac {V(x^2+y^2)\Sigma^2}{4\mu^2F^2})\\
&&\hspace*{-1cm}
\times \int_0^1 tdt e^{-2V \mu^2F^2t}(I_0^*(z\Sigma Vt) - 
\frac{I_0^*(z\Sigma V)I_0mV\Sigma t}{I_0(mV\Sigma)})I_0(z\Sigma Vt).\nn
\ee
It was first obtained from the random matrix
model  \cite{james}  using the method of complex orthogonal polynomials
developed in \cite{fyodorov-rev,Akemann}.
To appreciate this result, we considers its asymptotic expansion for 
$V\to \infty$.
For $m = 0$ and $x> 0$ we can derive the asymptotic result 
for the difference
\be
\rho^{\rm quen}(x,y,\mu)-\rho^{\rm full}(x,y,\mu) &\sim&  \frac{\sqrt{z\Sigma}}{2\mu^3F^3}
e^{\frac{Vz^{*\,2}\Sigma^2}{8\mu^2F^2} - \frac {V x^2 \Sigma^2}{2\mu^2}} 
e^{z^*\Sigma V}.
\label{fullasym}
\ee
The behavior near the extremum at $(x,y) =(4\mu^2F^2/3\Sigma,0) $ is
given by
\be
\rho^{\rm quen}(x,y,\mu)-\rho^{\rm full}(x,y,\mu)
 \sim e^{\frac 23 \mu^2 F^2 V} e^{\frac 23 iy \Sigma V}.
\ee
We have oscillations on the scale $ 1/\Sigma V$ with an amplitude that
diverges exponentially with the volume \cite{AOSV}. 
In the thermodynamic limit, these oscillations are visible
in a domain where the real part of the
exponent in (\ref{fullasym}) is positive. 
This region is given by intersection of the 
 inside of the ellipses
\be
3(x \pm \frac 43 \mu^2 F^2 \Sigma)^2 +y^2 = \frac {16}3 \mu^4 F^2 \Sigma^2
\ee
and the strip $|x| < 2F^2\mu^2/\Sigma$. At the mean field level this
can be reinterpreted as a region where Kaon condensation takes 
place \cite{KT,OSV}.
 
\section{Conclusions} 
\label{conclu}

The existence of two formulations of QCD at low energy, first as a
 microscopic theory of quarks and gluons and second as an effective 
theory of weakly interacting Goldstone bosons, imposes powerful constraints 
on either of the theories. The effective theory is completely determined
by the symmetries of the microscopic theory, and the mass dependence
of the effective theory imposes sum rules on the inverse Dirac
eigenvalues. In particular this means that any theory with the
same symmetry breaking pattern and a mass gap will be subject to the
same constraints. The simplest microscopic theory is chiral Random Matrix
Theory. 

However, more can be done than constraining the inverse Dirac eigenvalues
by sum rules. The key observation is that the generating function 
for the resolvent amounts to a
partition function with additional flavors with a mass $z$ 
equal to the value for which the resolvent
is calculated. Again we have a microscopic theory and an effective
theory with the same low energy limit. Because $z$ is a free parameter,
it can always be chosen such that the Compton wavelength of the 
corresponding Goldstone boson is much larger than the size of the 
box. In this region the $z$ dependence of the partition function is determined
by the mass term of the chiral Lagrangian which is a simple matrix
integral.

To obtain the Dirac spectrum we have to quench the determinant corresponding
to $z$. This can be done in two ways: by the replica trick or by the
supersymmetric method. Although, the supersymmetric method is 
straightforward, the {\em naive} replica trick is technically somewhat simpler.
The problem is that the {\em naive} replica trick does not give correct
nonperturbative results. One way out is if the dependence on
the number of replicas $n$ is known as an analytical function of $n$ around
$n=0$. It was shown by Kanzieper that the $n$-dependence can be
obtained from the solution of the Painlev$\acute {\rm e}$ equation. 
The other way out, which has been advocated in these lectures,
is if the partition function for $n=0$ is related
by a recursion relation to partition functions with a nonzero integer number
of flavors. The replica limit of this Toda lattice equation gives us
nonperturbative correlation functions.
This is an efficient formulation of the problem. The
structure of the final answer already has the factorized 
structure of the Toda lattice formulation. 
We could also say that the supersymmetric partition function connects
two semi-infinite hierarchies.

New results with the Toda lattice method were obtained for QCD at
nonzero chemical potential. In this case
the low-energy effective partition functions are also related by the
Toda lattice equation. This made it possible to express the
microscopic
spectral density as the product of the partition function
with one fermionic flavor and the partition function with one
bosonic flavor. This result has later been reproduced by RMT
with the method of orthogonal polynomials.

More surprisingly the Toda lattice method also gives the correct
result for QCD at nonzero chemical potential with dynamical
fermions. Because of the phase of the fermion determinant, 
a breakdown of this method for this case would not have been a surprise.
However, the concept of integrability that also reigns this case, is
so powerful that replica limit can be taken in exactly the same way
as in the quenched case. The result for the spectral density shows
oscillations on the scale of $1/V$ and an amplitude that diverges 
exponentially with $V$. This structure is necessary 
to obtain a nonzero chiral condensate in the chiral limit.

The Toda lattice method has been applied to quite a few cases
in the symmetry class $\beta = 2$ (see 
\cite{kanzieper,SplitVerb1,kanznh,SplitVerb2,AOSV,anders}).
Our conjecture is that
all microscopic correlation functions in this class can be
obtained from the replica limit of a Toda lattice equation.
A much tougher problem is analysis of the replica limit
for the other Dyson classes. We are not aware of any progress
on this problem and encourage the reader to confront this challenge.

\begin{acknowledgments}
The organizers of the 2004 Les Houches Summer School on Applications of
Random Matrices in Physics are thanked for giving me the opportunity 
to present these lectures as well as for providing a pleasant and
stimulating atmosphere.
G. Akemann, D. Dalmazi, P. Damgaard, A. Halasz, J.C. Osborn, 
M. Sener, L. Shifrin, K. Splittorff, 
M. Stephanov, 
D. Toublan, H.A. Weidenmueller, T. Wettig  and M.R. Zirnbauer are
thanked for sharing their insights which led to the joint publications
on which
these lectures are based. Kim Splittorff is thanked for a critical 
reading of the manuscript.

\end{acknowledgments}

\end{document}